\documentclass{article} 
\usepackage[preprint]{colm2026_conference}

\usepackage{microtype}
\usepackage{url}
\usepackage{booktabs}
\usepackage{multirow}
\usepackage{graphicx}
\graphicspath{{COLM/updated_images/}}
\usepackage{enumitem}
\usepackage{wrapfig}
\usepackage{subcaption}

\usepackage{latexsym}
\usepackage{amsmath, amssymb}
\usepackage{bbm}

\usepackage{lineno}

\definecolor{citecolor}{RGB}{33, 137, 126}
\definecolor{urlcolor}{RGB}{160, 89, 3}
\definecolor{linkcolor}{RGB}{215, 60, 87}
\usepackage[
    pagebackref=true,
    breaklinks=true,
    colorlinks=true,
    citecolor=citecolor,
    urlcolor=urlcolor,
    linkcolor=linkcolor
]{hyperref}

\title{Justified or Just Convincing? Error Verifiability as a Dimension of LLM Quality}

\usepackage{color-edits}
\usepackage[dvipsnames]{xcolor}
\definecolor{confNLL}{HTML}{1F77B4}
\definecolor{confVerb}{HTML}{FF7F0E}
\definecolor{confPtrue}{HTML}{2CA02C}

\addauthor{xz}{BrickRed}
\addauthor{kt}{DarkOrchid}
\addauthor{rf}{ForestGreen}
\addauthor{gs}{Maroon}
\addauthor{sw}{NavyBlue}

\newcommand{\vnorm}{$v_{\text{bal}}$}

\newcommand{\lm}{\pi}                  
\newcommand{\judge}{\mathcal{R}}       
\newcommand{\GT}{G}                    
\newcommand{\resp}{r}                  
\newcommand{\just}{j}                  
\newcommand{\ans}{a}                   
\newcommand{\question}{q}              

\newcommand{\yao}{y_0}                 
\newcommand{\yaj}{y_j}                 
\newcommand{\cao}{c_0}                 
\newcommand{\caj}{c_j}                 

\newcommand{\vFP}{v_{\text{FP}}}       
\newcommand{\vTN}{v_{\text{TN}}}       
\newcommand{\vFN}{v_{\text{FN}}}       
\newcommand{\vTP}{v_{\text{TP}}}       
\newcommand{\vnormeq}{V_{\text{bal}}}  

\author{
  Xiaoyuan Zhu$^{1}$\thanks{Correspondence: \texttt{\textcolor{urlcolor}{xzhu9839@usc.edu}}. Code: \textcolor{urlcolor}{\url{https://github.com/xyzhu123/Verifiability}}} \quad
  Kimberly Le Truong$^{2}$ \quad
  Riccardo Fogliato$^{3}$ \quad
  Gokul Swamy$^{2}$ \\
  \bfseries Weijian Zhang$^{4}$ \quad
  Minglai Yang$^{4}$ \quad
  Longtian Ye$^{4}$ \quad
  Bangya Liu$^{4}$ \quad
  Minghao Liu$^{4}$ \\
  \bfseries Andrew Ilyas$^{2}$ \quad
  Zhiwei Steven Wu$^{2}$ \\[0.5em]
  \mdseries
  $^{1}$University of Southern California \quad
  $^{2}$Carnegie Mellon University \\
  $^{3}$Microsoft Core AI \quad
  $^{4}$2077AI
}

\begin{document}

\ifcolmsubmission
\linenumbers
\fi

\maketitle

\begin{abstract}
As LLMs are deployed in high-stakes settings, users must judge the correctness of individual responses, often relying on model-generated justifications such as reasoning chains or explanations. Yet, no standard measure exists for whether these justifications help users distinguish correct answers from incorrect ones. We formalize this idea as \textit{error verifiability} and propose \vnorm{}, a balanced metric that measures whether justifications enable raters to accurately assess answer correctness, validated against human raters who show high agreement. We find that neither common approaches, such as post-training and model scaling, nor more targeted interventions recommended  improve verifiability. We introduce two methods that succeed at improving verifiability: reflect-and-rephrase (RR) for mathematical reasoning and oracle-rephrase (OR) for factual QA, both of which improve verifiability by incorporating domain-appropriate external information. Together, our results establish error verifiability as a distinct dimension of response quality that does not emerge from accuracy improvements alone and requires dedicated, domain-aware methods to address.
\end{abstract}

\section{Introduction}
Large language models (LLMs) are increasingly being deployed in high-stakes domains, where it is crucial for users to understand the justification behind an LLM's response. Within some domains, justifications produced by LLMs, whether in the form of a reasoning chain, explanation, or supporting arguments, have become treated as evidence that can support a medical diagnosis \citep{liu2025generalist, tu2024conversationaldiagnosticai}, legal decision \citep{zheng2025reasoning, fan2026lexam}, or scientific research \citep{gottweis2025aicoscientist, lu2024aiscientist}. Additionally, some have started to use LLM-generated natural-language explanations as an interpretability tool for model predictions and dataset patterns \citep{singh2024rethinking} and for describing models' own internal computations \citep{li2026training}. Despite this reliance on LLM-produced justifications, there are no standard measures of how capable a model is at producing justifications. Standard benchmarks only evaluate models by aggregate accuracy \citep{hendrycks2021mmlu, wang2024mmlupro, cobbe2021gsm8k, lightman2023verifystep, lin2022truthfulqa}. Furthermore, recent work suggests that these justifications are unreliable guides: explanations rarely enable users to verify AI predictions \citep{fok2023verifiability}, can inflate confidence in incorrect answers even when they add no informational value \citep{steyvers2025know}, may not faithfully reflect the model's actual reasoning process \citep{turpin2023unfaithful, lanham2023faithfulness}, and can be sycophantic, favoring responses that match user beliefs over truthful ones \citep{sharma2025sycophancy}. This body of work calls for better verification methods for LLM justifications \citep{barez2025chain}. Rather than how the justification may reflect the model's confidence or its true thinking process, we focus on how it impacts the downstream user.

We formalize this measure as \textit{error verifiability} (\S\ref{sec:methodology}), which assesses whether a reader can accurately determine whether a provided answer is correct given an LLM's justification. We consider a justification high quality if it improves a user's ability to discriminate correct from incorrect answers, and low quality if it misleads users into accepting incorrect answers or rejecting correct ones.
To operationalize error verifiability while isolating the effect of the justification, we propose \vnorm{}, a metric balanced across two conditions: the correctness of the model's response and a reader's baseline judgment without any justification.
We instantiate \vnorm{} using LLM-as-a-judge raters as a scalable approach to measuring verifiability. We conducted a human subject study to assess how human participants use justifications and found high agreement between human raters and LLMs. Further, we found that existing models lead to poor error verifiability across both LLM and human raters, demonstrating that verifiability is an ongoing, practical challenge (\S\ref{sec:inter_rater_agreement}).


Then, a natural question is whether existing approaches to improving model quality yield better justifications, and whether alternative methods can improve error verifiability. Across mathematical reasoning and factual knowledge QA benchmarks, we demonstrate that neither common approaches (post-training (\S\ref{sec:post_training}) and model scaling (\S\ref{sec:stronger_models})) nor more targeted interventions (stylistic rephrasing (\S\ref{sec:math_rephrase}) and calibrated linguistic confidence (\S\ref{sec:calibrated_confidence}) improve verifiability. 
We introduce two methods that succeed in improving verifiability: reflect-and-rephrase (RR) for mathematical reasoning (\S\ref{sec:math_rephrase}) and oracle-rephrase (OR) for factual knowledge QA (\S\ref{sec:factual_qa}). We present these methods as early demonstrations that verifiability can be improved through the incorporation of domain-appropriate external information. Because these methods are domain-specific and require some external information, extending them to broader settings and developing training-time approaches that directly optimize for verifiability remain open challenges. Across these analyses, we highlight three key findings:
\begin{itemize}[leftmargin=*]
    \item \textbf{Verifiability does not follow accuracy.} Post-training and model scaling greatly improve accuracy but do not change or worsen verifiability. This degradation is concentrated on wrong answers, exactly where verifiability matters most.
    \item \textbf{Surface-level modifications are insufficient.} Neither stylistic rephrasing nor calibrated linguistic confidence improves verifiability; changing how a justification is presented, without changing what information it contains, does not help raters catch errors.
    \item \textbf{Effective improvement requires domain-appropriate external information.} Both our proposed improvement methods go beyond information in the original justification. For math, RR cross-checks against alternative samples to surface inconsistencies that flag errors; for factual QA, OR uses external fact-checking to supply verification signals the model itself cannot provide.
\end{itemize}

Together, our results establish error verifiability as a distinct dimension of response quality:
\begin{itemize}[leftmargin=*]
  \item \textbf{We formalize error verifiability and propose \vnorm{},} a balanced metric that measures whether justifications help raters correctly judge the correctness of model answers.
  \item \textbf{We conduct a comprehensive evaluation,} showing that post-training, model scaling, stylistic rephrasing, and confidence calibration all fail to consistently improve verifiability.
  \item \textbf{We introduce two methods that improve verifiability} for mathematical reasoning and factual knowledge QA by injecting domain-appropriate external information.
\end{itemize}

\section{Related Works}
\paragraph{Overreliance on AI explanations.}
Several studies show that AI explanations increase user agreement with model outputs regardless of correctness \citep{bansal2021doesexceedpartseffect, kim2025fostering, steyvers2025know}, and that interventions such as cognitive forcing functions \citep{bucinca2021trust} or reliance calibration \citep{bo2025relyrelyevaluatinginterventions, schemmer2023appropriate} reduce overreliance without improving appropriate reliance. \citet{fok2023verifiability} argue that explanations rarely yield complementary human-AI performance because they do not support correctness verification, and \citet{ibrahim2025measuringmitigatingoverreliancenecessary} argue that RLHF amplifies overconfidence.
These works diagnose overreliance and test behavioral interventions but do not offer a formal metric for whether justifications help users distinguish correct from incorrect answers; we address this with \vnorm{} and show that improving verifiability requires incorporating domain-appropriate external information.

\paragraph{Simulatability of explanations.}
Simulatability asks whether an explanation helps users predict what a model will do \citep{hase2020evaluatingexplainableaialgorithmic, chen2023modelsexplainthemselvescounterfactual}. Recent work extends this notion to pragmatic perturbations \citep{hong2026llmselfexplanationshelpusers}, generation tasks \citep{limpijankit2025counterfactualsimulatabilityllmexplanations}, and training-time objectives \citep{hase2026counterfactualsimulationtrainingchainofthought}. \citet{mayne2026positivecasefaithfulnessllm} find that frontier models exhibit privileged self-knowledge that does not translate into verifiable justifications.
Simulatability evaluates whether explanations help predict model behavior under counterfactual inputs, whereas verifiability directly measures the trust decision users face in deployment: whether a justification helps judge the correctness of a given answer.

\paragraph{Detecting LLM errors.}
Prover-Verifier Games train models to produce legible reasoning that transfers to human verifiers \citep{kirchner2024proververifiergamesimprovelegibility, kim2026mitigatinglegibilitytaxdecoupled}. Process supervision \citep{lightman2023verifystep}, listener-aware fine-tuning \citep{stengeleskin2024lacielistenerawarefinetuningconfidence}, and self-verification pipelines \citep{dhuliawala2023chainofverificationreduceshallucinationlarge} each target error detection from a different angle. \citet{zhou2026interactive} show that presentation format alone affects error detection rates, and \citet{aggarwal2026evaluatingchainofthoughtreasoningreusability} find that CoT verifiability does not correlate with accuracy.
These approaches modify training procedures or model architectures to improve reasoning legibility. In contrast, we formalize verifiability as a post-hoc evaluation criterion independent of how the model was trained, and show that inference-time rephrasing with external information can improve it without retraining.

\section{Problem Formulation}\label{sec:preliminaries}
\paragraph{Setting.}
Justifications $\just$ generated by an LLM often serve as the primary signal available for determining if an LLM's answer $\ans$ to a question can be trusted. We evaluate whether the justification leads users to the correct verdict.
Let $\lm$ be an LLM.
Given a question $\question$, $\lm(\question)$ produces a response $\resp = (\just, \ans)$ sampled from a distribution induced by $\lm$. The justification, $\just$ is often represented by the reasoning chain or explanation displayed to the user preceding the LLM's answer.
$\just$ does not include any hidden chain-of-thought that the model may use internally but is not revealed to the user.
We assume there exists some ground truth answer and denote the correctness of $\ans$ by $\GT \in \{0,1\}$, where $\GT=1$ if and only if $\ans$ is correct.

\paragraph{Rater and evaluation settings.}
We consider two evaluation settings that differ in the information available to the rater. A \emph{rater} can be a third-party LLM or a human that receives information about a response and produces a binary judgment $y \in \{0,1\}$ indicating whether they believe $\ans$ is correct.
Under the Answer-Only (AO) setting, the rater receives $(\question, \ans)$ and outputs $\yao \in \{0,1\}$, representing verification based solely on the answer.
Under the Answer+Justification (AJ) setting, the rater receives $(\question, \just, \ans)$ and outputs $\yaj \in \{0,1\}$, representing verification with access to $\just$.
The specific rater configurations used for AO and AJ may vary by application; we discuss our choice and its implications in \S\ref{sec:inter_rater_agreement}.
Two special cases are worth noting. First, the same rater may serve in both settings; $\cao$ then reflects the rater's baseline judgment, and \vnorm{} measures whether the justification shifts that judgment toward or away from the correct verdict. Second, the AO and AJ raters may share the same model but differ in evaluation strategy (e.g., with or without reasoning tokens). This can be viewed as a single rater adopting different levels of deliberation rather than two distinct raters; the difference in strategy isolates the contribution of the justification from that of additional reasoning effort.

Let $\cao$ and $\caj$ denote the correctness in the AO and AJ settings respectively.
\[
  \cao = \mathbbm{1}[\yao = \GT], \qquad \caj = \mathbbm{1}[\yaj = \GT].
\]
$\cao = 1$ indicates that the AO setting yields a correct verdict, and $\caj = 0$ indicates that the AJ setting yields an incorrect verdict.

\paragraph{Why $\cao$ and $\caj$?}
One might evaluate justification quality by looking at $\caj$ alone, but $\caj$ does not indicate whether the correct verdict was driven by the justification or by the rater's own ability. A high $\caj$ may simply mean the answer was easy to verify regardless of the justification, while a low $\caj$ may reflect either a misleading justification or an inherently difficult instance.
Conditioning on $\cao$ partitions instances by their verification difficulty without access to the justification, so that a good metric must account for both cases: justifications that help on otherwise-hard instances ($\cao = 0$) and justifications that preserve correct outcomes on easy ones ($\cao = 1$).

\section{Measuring Verifiability with \vnorm{}}\label{sec:methodology}
\looseness=-1
We define \textit{verifiability} as the degree to which a justification $\just$ enables a rater to assess the correctness of $\ans$. A highly verifiable justification exposes errors when the answer is wrong and confirms correctness when it is right. We operationalize this definition through \vnorm{}.

\paragraph{Four verification scenarios.}
The pair $(\GT, \cao)$ partitions responses into four scenarios based on the ground-truth correctness and the baseline verification outcome. Interpreting the AO judgment as a binary classifier with label $\GT$ yields the scenarios: true positive (TP), true negative (TN), false positive (FP), and false negative (FN).
The role of the justification differs across these scenarios: in the FP and FN cells, the justification must provide enough signal to overcome the difficulty that the baseline setting reveals, while in the TP and TN cells, it must not degrade an otherwise correct verification outcome.


\paragraph{\vnorm{}: balanced verifiability.}
We define \vnorm{} (Balanced Verifiability) as the average correctness of the rater under AJ across the four scenarios:
\begin{align*}
  \vFP &= \mathbb{E}\!\left[\caj \;\middle|\; \GT=0,\;\cao=0\right], &
  \vTN &= \mathbb{E}\!\left[\caj \;\middle|\; \GT=0,\;\cao=1\right], \\
  \vFN &= \mathbb{E}\!\left[\caj \;\middle|\; \GT=1,\;\cao=0\right], &
  \vTP &= \mathbb{E}\!\left[\caj \;\middle|\; \GT=1,\;\cao=1\right],
\end{align*}
\[
  \vnormeq = \frac{1}{4}\!\left(\vFP + \vTN + \vFN + \vTP\right).
\]
By assigning equal weight to each cell, \vnorm{} is balanced on two conditions: the correctness of the model's response ($\GT$), so that a rater must genuinely detect errors rather than simply agree with a usually-correct model; and the baseline verification outcome ($\cao$), so that the metric disentangles the contribution of the justification from the baseline difficulty of each instance.
For example, if a model is correct on 90\% of questions, a rater who blindly accepts every answer already achieves 90\% overall accuracy, even without reading any justification. An unbalanced metric would assign this rater a high score despite its inability to detect errors. \vnorm{} avoids this by weighting the incorrect-answer cells equally with the correct-answer cells. Equal weighting ensures that performance on error detection and correctness confirmation, as well as on easy and hard instances, contributes equally to the final score.
A \vnorm{} of $0.5$ corresponds to random-level verification; a value of $1.0$ indicates that justifications are always associated with correct verification regardless of the baseline AO outcome.

\section{Experimental Design}
\paragraph{Tasks and benchmarks.}
We focus on two logical reasoning tasks: mathematical reasoning and factual knowledge QA.
For mathematical reasoning, we use GSM8K~\citep{cobbe2021gsm8k} and MATH500~\citep{lightman2023verifystep}; GSM8K contains grade-school arithmetic word problems, while MATH500 covers competition-level problems spanning algebra, geometry, and number theory, providing a range of difficulty levels. 
For factual knowledge QA, we use MMLU~\citep{hendrycks2021mmlu}, MMLU-Pro~\citep{wang2024mmlupro}, and TruthfulQA~\citep{lin2022truthfulqa}; MMLU and MMLU-Pro provide broad academic coverage at different difficulties, while TruthfulQA specifically targets questions where models tend to produce confident but incorrect answers, providing a natural stress test for verifiability. For each benchmark, we randomly sample 200 questions. Details on response generation, ground-truth labeling, and data processing are provided in Appendix~\ref{app:data_preparation}.


\paragraph{LLM-as-a-judge raters.}
Since collecting human judgments at scale is infeasible, we use LLM-as-a-judge as a surrogate for the rater $\judge$.
We employ three models from distinct provider families: GPT-4.1-mini, Claude-Haiku-4.5, and Gemini-2.5-Flash-Lite.
All raters operate with temperature $0.0$. 
We evaluate two rater modes in single-turn evaluation: \emph{direct} with no thinking tokens allocated, and \emph{thinking}, which allocates a 256-token scratchpad before a forced response.
Rater configurations and prompts are provided in Appendix~\ref{app:judge_prompts}. Full model names and API identifiers for all models used in this work are listed in Appendix~\ref{app:model_names}.

\section{Validating LLM-as-a-Judge with a Human Study}\label{sec:inter_rater_agreement}
We validate LLM-as-a-judge as a surrogate for human raters with a human study comparing LLM and human judgments on a sample of MATH500 responses. We use this study to determine which rater configurations most closely approximate human behavior for computing \vnorm{}. The full implementation details are in Appendix~\ref{app:human_study}.

\paragraph{Setup.}
We sample 40 questions and their corresponding responses from MATH500 at difficulty levels 3--4, corresponding to American Mathematics Competition (AMC) 10/12 level, stratified into 10 items per verification scenario using direct AO labels.
We recruit 19 undergraduate students who passed a placement test on mathematical and English reading skills. Each participant rates 8 AO and 8 AJ items, interleaved to prevent ordering effects; each item appears in at most one setting per participant.
Three LLM raters (GPT-4.1-mini, Claude-Haiku-4.5, and Gemini-2.5-Flash-Lite) evaluate all 40 items under four settings: \textbf{AO}, \textbf{AJ}, and their thinking-mode counterparts \textbf{AO-CoT} and \textbf{AJ-CoT}.
We measure agreement with Cohen's~$\kappa$ \citep{Cohen1960ACO} on all (LLM, human) judgment pairs.

\subsection{Results}

\begin{wraptable}{r}{0.50\textwidth}
\vspace{-26pt}
\centering
\caption{LLM--human $\kappa$ and accuracy.}
\vspace{-3pt}
\label{tab:human_study}
\footnotesize
\setlength{\tabcolsep}{4pt}
\begin{tabular}{lcccc}
\toprule
& \textbf{AO} & \textbf{AO-CoT} & \textbf{AJ} & \textbf{AJ-CoT} \\
\midrule
$\kappa$     & 0.065 & 0.481 & 0.501 & 0.488 \\
Acc (LLMs)   & 0.550 & 0.850 & 0.848 & 0.908 \\
Acc (Human)  & 0.836 & ---   & 0.809 & ---   \\
\bottomrule
\end{tabular}
\vspace{-16pt}
\end{wraptable}

Table~\ref{tab:human_study} reports accuracy and LLM--human $\kappa$ for all four settings, averaged across LLM raters. We focus on $\kappa$ as the primary alignment measure; full per-rater results are in Appendix~\ref{app:inter_rater_results}.

\paragraph{LLM raters approximate participants when capability is matched.}
In the AO setting, direct LLM judgments show near-zero agreement with participants ($\kappa = 0.065$) and much lower accuracy ($0.550$ vs.\ $0.836$), indicating a capability mismatch. Adding reasoning (AO-CoT) raises both accuracy ($0.850$) and agreement ($\kappa = 0.481$), bringing LLM behavior closer to that of participants.
In the AJ setting, direct AJ and AJ-CoT achieve similar agreement ($\kappa = 0.501$ vs.\ $0.488$); unlike AO, adding reasoning does not meaningfully increase alignment.

\paragraph{Justifications do not always help and can hurt.}
For participants, who already perform well under AO ($0.836$), justifications slightly decrease accuracy to $0.809$, adding little signal beyond what they can assess independently.
In contrast, LLM raters in the direct AO setting achieve only $0.550$, but direct AJ raises accuracy to $0.848$, showing that justifications help substantially when the baseline is weaker.
These findings imply that justifications are most useful when they provide information the rater would not otherwise have, motivating conditioning on $\cao$ when measuring verifiability.

\paragraph{Informing the evaluation protocol.}
Based on these results, we adopt \textbf{AO-CoT} for the AO setting and \textbf{direct AJ} for the AJ setting in all subsequent experiments, as each most closely matches participant behavior ($\kappa = 0.481$ and $0.501$).
Under this protocol, $\cao$ from AO-CoT partitions instances by verification difficulty from the answer alone, and $\caj$ from direct AJ measures outcomes when the justification is available (\S\ref{sec:preliminaries}). \vnorm{} then captures whether justifications are associated with better or worse outcomes across this partition.
\vnorm{} is compatible with any (AO, AJ) configuration; practitioners may choose other settings depending on the use case.

\section{Does Improving Model Capability Improve Verifiability?}
\label{sec:modelCap}

We examine whether standard approaches to improving model capability, namely post-training (\S\ref{sec:post_training}) and model scaling (\S\ref{sec:stronger_models}), also improve verifiability.

\subsection{Post-Training Does Not Consistently Improve Verifiability}\label{sec:post_training}
\begin{figure}[t]
\centering
\includegraphics[width=\textwidth]{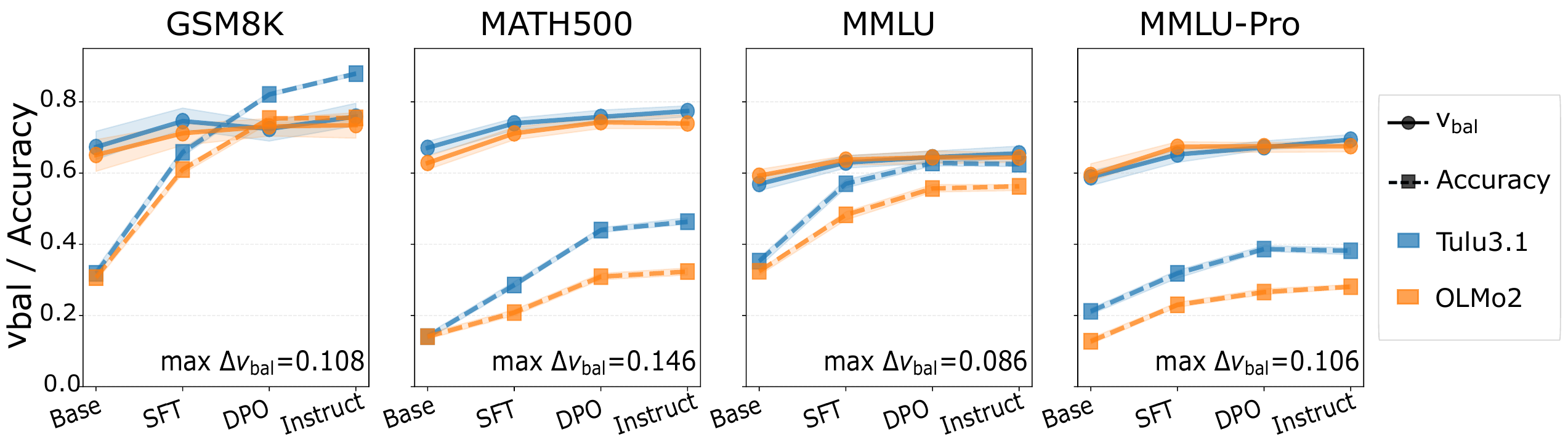}
\caption{\vnorm{} and accuracy across post-training checkpoints for Tulu3.1-8B and OLMo2-7B.}
\label{fig:checkpoint_vnorm}
\end{figure}

\paragraph{Setup.}
To isolate the effect of post-training on verifiability, we track two open-weight model families, Tulu3.1-8B and OLMo2-7B, at four stages of their training pipeline: the pre-trained \textbf{Base} checkpoint, supervised fine-tuning (\textbf{SFT}), direct preference optimization (\textbf{DPO}), and the final released \textbf{Instruct} model.
For each (checkpoint, benchmark) pair, we report \vnorm{} averaged over the three LLM raters. Full results are in Appendix~\ref{app:checkpoint_full}.

\paragraph{Accuracy improves substantially; verifiability remains stagnant.}
Figure~\ref{fig:checkpoint_vnorm} plots accuracy and \vnorm{} across four post-training stages.
Accuracy rises from Base to Instruct (up to $+0.56$ on GSM8K for Tulu3.1-8B).
\vnorm{}, by contrast, changes little throughout post-training and is non-monotonic: on GSM8K, for instance, \vnorm{} peaks at SFT and then declines after DPO.

\paragraph{Preference optimization masks errors harder to detect.}
A per-cell analysis (Appendix~\ref{app:checkpoint_full}) reveals that the drop in \vnorm{} between SFT and DPO is concentrated on the incorrect answers (FP and TN cells). On GSM8K, for example, the FP score of Tulu3.1-8B drops from $0.717$ at SFT to $0.602$ after DPO.
This suggests that by improving the surface fluency of all responses, preference optimization makes the justifications accompanying incorrect answers look more convincing, thereby removing cues that a rater could otherwise use to flag errors.

\subsection{Stronger Models Do Not Always Have Better Verifiability}\label{sec:stronger_models}
\begin{figure}[t]
\centering
\includegraphics[width=\textwidth]{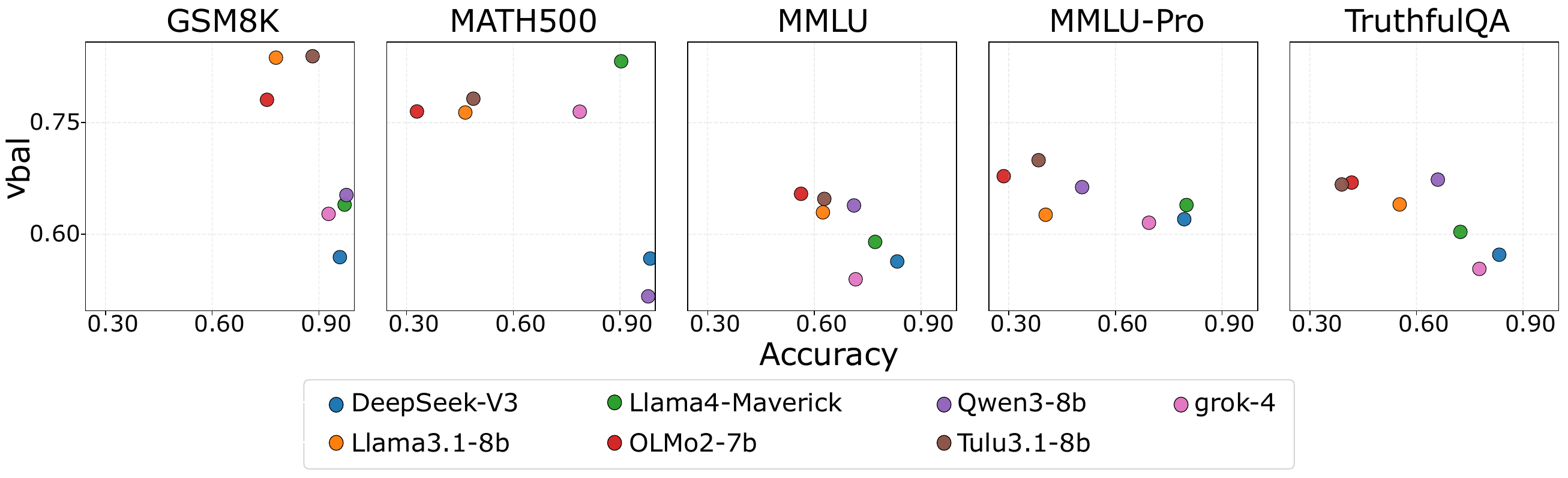}
\caption{Accuracy vs.\ \vnorm{} across models and datasets.}
\label{fig:acc_vs_vnorm}
\end{figure}

\paragraph{Setup.}
We extend the analysis beyond single model families to compare seven models across different scales and providers: three open-weight 7--8B models (Llama3.1-8B, Tulu3.1-8B, OLMo2-7B) and four frontier models (Qwen3, DeepSeek-V3, Llama4-Maverick, Grok-4).
Each model is evaluated on all five benchmarks, with \vnorm{} averaged over the three LLM raters. See a detailed breakdown of the results in Appendix~\ref{app:stronger_models_full}.

\paragraph{Higher accuracy does not result in higher verifiability.}
Figure~\ref{fig:acc_vs_vnorm} plots accuracy against \vnorm{} for every (model, dataset) pair.
If stronger models produced more verifiable justifications, we would expect models with higher accuracy to also achieve higher \vnorm{}. Instead, the opposite trend holds consistently across all five benchmarks: the most accurate models tend to rank among the lowest in \vnorm{}, while weaker models such as OLMo2-7B often achieve the highest \vnorm{}. This pattern persists on both mathematical reasoning and factual knowledge benchmarks.

\paragraph{Stronger models lose verifiability on incorrect answers.}
The per-cell breakdown (Appendix~\ref{app:stronger_models_full}) shows that the \vnorm{} gap between stronger and weaker models is concentrated in the FP and TN cells, both involving incorrect answers. On GSM8K, for example, frontier models such as Qwen3 and Llama4-Maverick see large drops in FP and TN scores relative to smaller models, while TP scores remain comparable across all models. The same pattern holds on MATH500 and MMLU.
Verifiability consistency also varies across models: OLMo2-7B maintains relatively stable \vnorm{} across benchmarks, while Qwen3 performs well on some but poorly on others despite similar accuracy levels.

The disconnect between accuracy and verifiability highlights substantial room to improve verifiability without compromising performance. Moreover, the tendency of post-training and larger models to reduce verifiability raises questions about the effectiveness of current training approaches.

\section{Improving Verifiability}
Our findings in \S\ref{sec:modelCap} motivate more targeted interventions. We consider rephrasing methods, a low-cost inference-time approach to improving LLM responses~\citep{madaan2023selfrefine, deng2024rephraserespond, ning2024skeletonofthought, shu2023rewritelm}, as well as confidence-focused rephrasing including linguistic calibration (\S\ref{sec:calibrated_confidence}) and selective rephrasing based on confidence (Appendix~\ref{app:best_of_n}). However, all these methods fail to improve verifiability. We find that effective interventions must account for domain-specific needs and propose separate rephrasing techniques for mathematical benchmarks (\S\ref{sec:math_rephrase}) and factual knowledge QA (\S\ref{sec:factual_qa}).

This setting is formalized as follows: given a response $(a, j)$, we produce $(a, j')$ where the final answer $a$ is preserved and only the justification $j$ is modified, with the goal of improving verifiability without altering model predictions.

\subsection{Mathematical Benchmarks}
\label{sec:math_rephrase}

\paragraph{Stylistic rephrase baselines.}
Prior work shows that reducing the difficulty of LLM explanations reduces overreliance on incorrect predictions~\citep{vasconcelos2023explanations} and that restructuring explanations into more readable formats improves human error detection~\citep{zhou2026interactive}.
Following these findings, we design three baseline rephrasing methods that modify justification style to reduce cognitive effort for raters: \textit{Structured} (\textsc{Struct.}) reorganizes into a step-by-step numbered format, \textit{Professional} (\textsc{Prof.}) improves structural clarity and term consistency through formal rewriting, and \textit{Simplified} (\textsc{Simpl.}) simplifies language and removes redundancy. Full prompts are in Appendix~\ref{app:rephrase_prompts}.

\paragraph{Reflect-and-rephrase.}
We additionally introduce \textit{reflect-and-rephrase} (\textsc{RR}), which builds on the hypothesis that inconsistencies across multiple responses reveal potential errors. The method proceeds in two steps:
\begin{itemize}[noitemsep, topsep=2pt]
    \item \textit{Reflect}: a rephrase model analyzes the target response against $k$ alternative responses, producing a reflection on where the responses agree or diverge.
    \item \textit{Rephrase}: the rephrase model rewrites the justification conditioned on this reflection, surfacing inconsistencies as explicit uncertainty markers.
\end{itemize}
Prompts are in Appendix~\ref{app:rr_prompts}.
\begin{table}[t]
\centering
\caption{\vnorm{} across all benchmarks ($\Delta$ vs.\ Base). Per-cell results in Appendix~\ref{app:rephrase_full}.}
\label{tab:rephrase}
\small
\setlength{\tabcolsep}{4pt}
\begin{tabular}{@{}llrr@{\hspace{1cm}}rr@{}}
\toprule
\textbf{Model} & & \textbf{MATH500} & \textbf{GSM8K} & \textbf{MMLU} & \textbf{TruthfulQA} \\
\midrule
\multirow{6}{*}{Llama3.1-8B}
 & Base            & 0.763 & 0.773 & 0.625 & 0.636 \\
 & \textsc{Prof.} $\Delta$   & $+$0.028 & $-$0.040 & $+$0.029 & $+$0.009 \\
 & \textsc{Struct.} $\Delta$ & $+$0.012 & $-$0.047 & $+$0.037 & $+$0.022 \\
 & \textsc{Simpl.} $\Delta$  & $+$0.002 & $-$0.080 & $+$0.043 & $+$0.026 \\
 & \textbf{RR $\Delta$ }     & $\mathbf{+0.040}$ & $\mathbf{+0.027}$ & $+$0.036 & $+$0.007 \\
 & \textbf{OR $\Delta$}      & --- & --- & $\mathbf{+0.086}$ & $\mathbf{+0.069}$ \\
\midrule
\multirow{6}{*}{Tulu3.1-8B}
 & Base            & 0.778 & 0.757 & 0.657 & 0.685 \\
 & \textsc{Prof.} $\Delta$   & $-$0.003 & $-$0.023 & $+$0.009 & $-$0.004 \\
 & \textsc{Struct.}. $\Delta$ & $-$0.010 & $-$0.030 & $+$0.002 & $+$0.013 \\
 & \textsc{Simpl.} $\Delta$  & $-$0.017 & $-$0.059 & $+$0.000 & $+$0.002 \\
 & \textbf{RR $\Delta$ }     & $\mathbf{+0.024}$ & $\mathbf{+0.048}$ & $+$0.000 & $-$0.069 \\
 & \textbf{OR $\Delta$ }     & --- & --- & $\mathbf{+0.066}$ & $\mathbf{+0.038}$ \\
\midrule
\multirow{6}{*}{OLMo2-7B}
 & Base            & 0.734 & 0.736 & 0.640 & 0.670 \\
 & \textsc{Prof.} $\Delta$   & $+$0.026 & $-$0.015 & $+$0.019 & $-$0.011 \\
 & \textsc{Struct.} $\Delta$ & $+$0.030 & $-$0.017 & $+$0.013 & $-$0.014 \\
 & \textsc{Simpl.} $\Delta$  & $+$0.020 & $-$0.058 & $+$0.002 & $-$0.011 \\
 & \textbf{RR $\Delta$}      & $\mathbf{+0.061}$ & $\mathbf{+0.083}$ & $+$.006 & $-$.027 \\
 & \textbf{OR $\Delta$}      & --- & --- & $\mathbf{+0.045}$ & $\mathbf{+0.026}$ \\
\bottomrule
\end{tabular}
\end{table}
Unless stated otherwise, we use Tulu3.1-8B to rephrase.

\paragraph{Results.}
Stylistic rephrasing methods, which aim to reduce cognitive effort without altering content, yield inconsistent \vnorm{} changes (Table~\ref{tab:rephrase}). All baselines decrease \vnorm{} for every model on GSM8k (up to $-$0.080), with small, mixed effects on MATH500. 
In contrast, \textsc{RR} produces consistent gains across all (model, dataset) pairs, with improvements concentrated in the FP and TN cells (Appendix~\ref{app:rephrase_full}), suggesting that \textsc{RR} primarily helps raters identify incorrect responses, without deteriorating verifiability for the correct answers.


\subsection{Factual Knowledge QA}
\label{sec:factual_qa}

\paragraph{Verifiability improvement methods do not transfer across domains.}
While \textsc{RR} consistently improves verifiability on mathematical benchmarks, this does not transfer to factual knowledge QA (Table~\ref{tab:rephrase}). Neither the baselines nor \textsc{RR} yield consistent \vnorm{} gains across models and datasets. We hypothesize that this discrepancy arises from the nature of the justifications. Mathematical reasoning involves calculations and logical deductions whose correctness can be checked for inconsistencies, providing \textsc{RR} with a reliable error signal. In contrast, factual QA justifications decompose into individual factual claims, each still requiring verification. Without sufficient information to verify these claims, the model cannot reliably detect errors.

\paragraph{Oracle fact checking enables verifiability.}
To test our hypothesis that factual QA requires external verification signals, we investigate whether supplying the rephrasing model with external information to verify individual factual claims provides as the missing signal.
We implement this idea with \textit{oracle rephrase} (\textsc{OR}): given a justification, we extract its atomic claims, verify each against an oracle model, judged as \textsc{Correct}, \textsc{Incorrect}, or \textsc{Not\_Verifiable}, and rewrite the justification with explicit inline annotations for any flagged claims. We use a strong model, Claude-Sonnet-4.5, as the fact checker. Full experimental details are in Appendix~\ref{app:or_prompts}.

\textsc{OR} yields consistent \vnorm{} improvements across both datasets and all three models (Table~\ref{tab:rephrase}). This supports our hypothesis that factual errors invisible to the rephrasing model become detectable once oracle verification is provided, and that external verification is necessary for factual QA tasks. Further, different domains, such as mathematics and factual QA require different approaches to improve verifiability.

\subsection{Calibrated Confidence Does Not Improve Verifiability}
\label{sec:calibrated_confidence}
Beyond stylistic rephrasing, we also consider calibrating the \textit{linguistic confidence} expressed by a response to match the model's internal certainty \citep{lin2022teaching, xiong2024llmsuncertainty}. This approach draws from prior work suggesting that if the model is wrong, rephrasing the response to sound uncertain may help raters discount it. However, we find that LLMs often do not know when they are wrong, making calibrating linguistic confidence not a useful method. The full rephrase prompt is provided in Appendix~\ref{app:uncertain_prompt}.

\paragraph{Internal confidence measures.}
We consider three measures of model confidence: \textit{NLL}, the average negative log-likelihood of the response tokens; \textit{Verbalized}, a self-reported confidence score on $[0, 1]$ elicited from the model after generating the response \citep{yang2024verbalizedconfidence}; and \textit{P(true)}, the probability the model assigns to ``True'' when prompted with ``Is the above answer correct?'' about its own response \citep{kadavath2022language}.

\paragraph{Procedure.}
For each measure, we rank responses by confidence and rephrase the bottom-$k\%$ to express uncertainty, while preserving the original information and conclusion. We sweep $k$ from $0\%$ to $100\%$ and measure \vnorm{} at each threshold.
If calibrated confidence were useful, the optimal $k$ should fall at an intermediate value, producing a \vnorm{} peak where only truly uncertain responses are hedged. Of course, this relies on the assumption that an LLM should be uncertain when its answer is incorrect.

\paragraph{Results.}
The optimal $k$ does not cluster at intermediate values (Figure~\ref{fig:calibrated_confidence}). In most settings, the peak falls at an extreme ($k{=}100\%$ or $k{=}0\%$), and the absolute \vnorm{} improvement over the baseline is small (typically below $0.02$).
This means that uniformly rephrasing all responses to sound uncertain is at least as effective as selectively targeting the least-confident ones. Current internal confidence measures lack the discriminative signal to identify which responses would benefit from hedging: the na\"ive strategy of hedging every justification already captures the limited gain that uncertain phrasing provides.

\section{Conclusion}
This work formalizes \textit{error verifiability} and introduces \vnorm{}, a balanced metric that measures whether LLM-generated justifications help raters judge answer correctness. Across seven models and five benchmarks, we find that post-training and model scaling yield large accuracy gains but leave verifiability flat or worse, particularly on incorrect answers. Stylistic rephrasing and calibrated linguistic confidence are similarly ineffective: changing presentation without changing content does not help raters detect errors. In contrast, our proposed methods, reflect-and-rephrase (RR) and oracle-rephrase (OR), consistently improve \vnorm{} by supplying domain-appropriate external information. A human study validates our LLM-as-a-judge protocol and confirms that the verifiability gap is a practical challenge. These results suggest that current training paradigms produce models that grow more persuasive as they become more capable, without a corresponding increase in error detectability. As LLMs are deployed in high-stakes domains, verifiability should be treated as a first-class evaluation criterion alongside accuracy.

\paragraph{Limitations and future work.} While our study establishes key insights, several limitations highlight promising directions for future research. First, our improvement methods are domain-specific. Developing unified approaches that generalize across domains remains an open challenge. Second, although our human study supports the LLM-as-a-judge protocol, the scale of human evaluation is limited; larger evaluations across diverse user populations would strengthen confidence in these findings. Such evaluations could also enable enable controlled analysis of how verifiability depends jointly on justification content and model capability (e.g., by comparing identical justifications across models). Third, our evaluation focuses on mathematical reasoning and factual knowledge QA, but other settings, such as code generation, medical QA, and legal reasoning, may present different challenges and methodologies. Finally, all proposed methods operate at inference time via post-hoc rephrasing; developing training-time objectives that directly optimize for verifiability, and extending these methods to open-domain settings where reliable ground truth is unavailable, are important future directions.

\clearpage
\bibliography{COLM/colm2026_conference}

@article{barez2025chain,
  title={Chain-of-Thought Is Not Explainability},
  author={Barez, Fazl and Wu, Tung-Yu and Arcuschin, Iv{\'a}n and Lan, Michael and Wang, Vincent and Siegel, Noah and Collignon, Nicolas and Neo, Clement and Lee, Isabelle and Paren, Alasdair and Bibi, Adel and Trager, Robert and Fornasiere, Damiano and Yan, John and Elazar, Yanai and Bengio, Yoshua},
  journal={Preprint, alphaXiv},
  year={2025},
  url={https://www.alphaxiv.org/abs/2025.02v1}
}

@misc{sharma2025sycophancy,
      title={Towards Understanding Sycophancy in Language Models}, 
      author={Mrinank Sharma and Meg Tong and Tomasz Korbak and David Duvenaud and Amanda Askell and Samuel R. Bowman and Newton Cheng and Esin Durmus and Zac Hatfield-Dodds and Scott R. Johnston and Shauna Kravec and Timothy Maxwell and Sam McCandlish and Kamal Ndousse and Oliver Rausch and Nicholas Schiefer and Da Yan and Miranda Zhang and Ethan Perez},
      year={2025},
      eprint={2310.13548},
      archivePrefix={arXiv},
      primaryClass={cs.CL},
      url={https://arxiv.org/abs/2310.13548}, 
}

@misc{lanham2023faithfulness,
      title={Measuring Faithfulness in Chain-of-Thought Reasoning}, 
      author={Tamera Lanham and Anna Chen and Ansh Radhakrishnan and Benoit Steiner and Carson Denison and Danny Hernandez and Dustin Li and Esin Durmus and Evan Hubinger and Jackson Kernion and Kamilė Lukošiūtė and Karina Nguyen and Newton Cheng and Nicholas Joseph and Nicholas Schiefer and Oliver Rausch and Robin Larson and Sam McCandlish and Sandipan Kundu and Saurav Kadavath and Shannon Yang and Thomas Henighan and Timothy Maxwell and Timothy Telleen-Lawton and Tristan Hume and Zac Hatfield-Dodds and Jared Kaplan and Jan Brauner and Samuel R. Bowman and Ethan Perez},
      year={2023},
      eprint={2307.13702},
      archivePrefix={arXiv},
      primaryClass={cs.AI},
      url={https://arxiv.org/abs/2307.13702}, 
}

@misc{turpin2023unfaithful,
      title={Language Models Don't Always Say What They Think: Unfaithful Explanations in Chain-of-Thought Prompting}, 
      author={Miles Turpin and Julian Michael and Ethan Perez and Samuel R. Bowman},
      year={2023},
      eprint={2305.04388},
      archivePrefix={arXiv},
      primaryClass={cs.CL},
      url={https://arxiv.org/abs/2305.04388}, 
}

@misc{lightman2023verifystep,
      title={Let's Verify Step by Step}, 
      author={Hunter Lightman and Vineet Kosaraju and Yura Burda and Harri Edwards and Bowen Baker and Teddy Lee and Jan Leike and John Schulman and Ilya Sutskever and Karl Cobbe},
      year={2023},
      eprint={2305.20050},
      archivePrefix={arXiv},
      primaryClass={cs.LG},
      url={https://arxiv.org/abs/2305.20050}, 
}

@misc{wang2024mmlupro,
      title={MMLU-Pro: A More Robust and Challenging Multi-Task Language Understanding Benchmark}, 
      author={Yubo Wang and Xueguang Ma and Ge Zhang and Yuansheng Ni and Abhranil Chandra and Shiguang Guo and Weiming Ren and Aaran Arulraj and Xuan He and Ziyan Jiang and Tianle Li and Max Ku and Kai Wang and Alex Zhuang and Rongqi Fan and Xiang Yue and Wenhu Chen},
      year={2024},
      eprint={2406.01574},
      archivePrefix={arXiv},
      primaryClass={cs.CL},
      url={https://arxiv.org/abs/2406.01574}, 
}

@misc{lin2022truthfulqa,
      title={TruthfulQA: Measuring How Models Mimic Human Falsehoods}, 
      author={Stephanie Lin and Jacob Hilton and Owain Evans},
      year={2022},
      eprint={2109.07958},
      archivePrefix={arXiv},
      primaryClass={cs.CL},
      url={https://arxiv.org/abs/2109.07958}, 
}

@misc{cobbe2021gsm8k,
      title={Training Verifiers to Solve Math Word Problems}, 
      author={Karl Cobbe and Vineet Kosaraju and Mohammad Bavarian and Mark Chen and Heewoo Jun and Lukasz Kaiser and Matthias Plappert and Jerry Tworek and Jacob Hilton and Reiichiro Nakano and Christopher Hesse and John Schulman},
      year={2021},
      eprint={2110.14168},
      archivePrefix={arXiv},
      primaryClass={cs.LG},
      url={https://arxiv.org/abs/2110.14168}, 
}

@misc{hendrycks2021mmlu,
      title={Measuring Massive Multitask Language Understanding}, 
      author={Dan Hendrycks and Collin Burns and Steven Basart and Andy Zou and Mantas Mazeika and Dawn Song and Jacob Steinhardt},
      year={2021},
      eprint={2009.03300},
      archivePrefix={arXiv},
      primaryClass={cs.CY},
      url={https://arxiv.org/abs/2009.03300}, 
}

@misc{lu2024aiscientist,
      title={The AI Scientist: Towards Fully Automated Open-Ended Scientific Discovery}, 
      author={Chris Lu and Cong Lu and Robert Tjarko Lange and Jakob Foerster and Jeff Clune and David Ha},
      year={2024},
      eprint={2408.06292},
      archivePrefix={arXiv},
      primaryClass={cs.AI},
      url={https://arxiv.org/abs/2408.06292}, 
}

@misc{gottweis2025aicoscientist,
      title={Towards an AI co-scientist}, 
      author={Juraj Gottweis and Wei-Hung Weng and Alexander Daryin and Tao Tu and Anil Palepu and Petar Sirkovic and Artiom Myaskovsky and Felix Weissenberger and Keran Rong and Ryutaro Tanno and Khaled Saab and Dan Popovici and Jacob Blum and Fan Zhang and Katherine Chou and Avinatan Hassidim and Burak Gokturk and Amin Vahdat and Pushmeet Kohli and Yossi Matias and Andrew Carroll and Kavita Kulkarni and Nenad Tomasev and Yuan Guan and Vikram Dhillon and Eeshit Dhaval Vaishnav and Byron Lee and Tiago R D Costa and José R Penadés and Gary Peltz and Yunhan Xu and Annalisa Pawlosky and Alan Karthikesalingam and Vivek Natarajan},
      year={2025},
      eprint={2502.18864},
      archivePrefix={arXiv},
      primaryClass={cs.AI},
      url={https://arxiv.org/abs/2502.18864}, 
}

@misc{fan2026lexam,
      title={LEXam: Benchmarking Legal Reasoning on 340 Law Exams}, 
      author={Yu Fan and Jingwei Ni and Jakob Merane and Yang Tian and Yoan Hermstrüwer and Yinya Huang and Mubashara Akhtar and Etienne Salimbeni and Florian Geering and Oliver Dreyer and Daniel Brunner and Markus Leippold and Mrinmaya Sachan and Alexander Stremitzer and Christoph Engel and Elliott Ash and Joel Niklaus},
      year={2026},
      eprint={2505.12864},
      archivePrefix={arXiv},
      primaryClass={cs.CL},
      url={https://arxiv.org/abs/2505.12864}, 
}

@inproceedings{zheng2025reasoning,
   series={CSLAW ’25},
   title={A Reasoning-Focused Legal Retrieval Benchmark},
   url={http://dx.doi.org/10.1145/3709025.3712219},
   DOI={10.1145/3709025.3712219},
   booktitle={Proceedings of the Symposium on Computer Science and Law on ZZZ},
   publisher={ACM},
   author={Zheng, Lucia and Guha, Neel and Arifov, Javokhir and Zhang, Sarah and Skreta, Michal and Manning, Christopher D. and Henderson, Peter and Ho, Daniel E.},
   year={2025},
   month=mar, pages={169–193},
   collection={CSLAW ’25} }

@misc{tu2024conversationaldiagnosticai,
      title={Towards Conversational Diagnostic AI}, 
      author={Tao Tu and Anil Palepu and Mike Schaekermann and Khaled Saab and Jan Freyberg and Ryutaro Tanno and Amy Wang and Brenna Li and Mohamed Amin and Nenad Tomasev and Shekoofeh Azizi and Karan Singhal and Yong Cheng and Le Hou and Albert Webson and Kavita Kulkarni and S Sara Mahdavi and Christopher Semturs and Juraj Gottweis and Joelle Barral and Katherine Chou and Greg S Corrado and Yossi Matias and Alan Karthikesalingam and Vivek Natarajan},
      year={2024},
      eprint={2401.05654},
      archivePrefix={arXiv},
      primaryClass={cs.AI},
      url={https://arxiv.org/abs/2401.05654}, 
}

@article{liu2025generalist,
  title={A generalist medical language model for disease diagnosis assistance},
  author={Xiaohong Liu and Hao Liu and Guoxing Yang and Zeyu Jiang and Shuguang Cui and Zhaoze Zhang and Huan Wang and Liyuan Tao and Yongchang Sun and Zhu Song and Tianpei Hong and Jin Yang and Tianrun Gao and Jiangjiang Zhang and Xiaohu Li and Jing Zhang and Ye Sang and Zhao Yang and Kanmin Xue and Song Wu and Ping Zhang and Jian Yang and Chunli Song and Guangyu Wang},
  journal={Nature Medicine},
  year={2025},
  volume={31},
  pages={932 - 942},
  url={https://api.semanticscholar.org/CorpusID:275425003}
}

@article{fok2023verifiability,
      title={In Search of Verifiability: Explanations Rarely Enable Complementary Performance in AI-Advised Decision Making}, 
      author={Raymond Fok and Daniel S. Weld},
      year={2024},
      eprint={2305.07722},
      archivePrefix={arXiv},
      primaryClass={cs.AI},
      url={https://arxiv.org/abs/2305.07722}, 
}

@article{steyvers2025know,
   title={What large language models know and what people think they know},
   volume={7},
   ISSN={2522-5839},
   url={http://dx.doi.org/10.1038/s42256-024-00976-7},
   DOI={10.1038/s42256-024-00976-7},
   number={2},
   journal={Nature Machine Intelligence},
   publisher={Springer Science and Business Media LLC},
   author={Steyvers, Mark and Tejeda, Heliodoro and Kumar, Aakriti and Belem, Catarina and Karny, Sheer and Hu, Xinyue and Mayer, Lukas W. and Smyth, Padhraic},
   year={2025},
   month=jan, pages={221–231} }

@article{vasconcelos2023explanations,
      title={Explanations Can Reduce Overreliance on AI Systems During Decision-Making}, 
      author={Helena Vasconcelos and Matthew Jörke and Madeleine Grunde-McLaughlin and Tobias Gerstenberg and Michael Bernstein and Ranjay Krishna},
      year={2023},
      eprint={2212.06823},
      archivePrefix={arXiv},
      primaryClass={cs.HC},
      url={https://arxiv.org/abs/2212.06823}, 
}

@inproceedings{zhou2026interactive,
      title={Improving Human Verification of LLM Reasoning through Interactive Explanation Interfaces}, 
      author={Runtao Zhou and Giang Nguyen and Nikita Kharya and Anh Totti Nguyen and Chirag Agarwal},
      year={2026},
      eprint={2510.22922},
      archivePrefix={arXiv},
      primaryClass={cs.HC},
      url={https://arxiv.org/abs/2510.22922}, 
}

@article{singh2024rethinking,
      title={Rethinking Interpretability in the Era of Large Language Models}, 
      author={Chandan Singh and Jeevana Priya Inala and Michel Galley and Rich Caruana and Jianfeng Gao},
      year={2024},
      eprint={2402.01761},
      archivePrefix={arXiv},
      primaryClass={cs.CL},
      url={https://arxiv.org/abs/2402.01761}, 
}

@article{li2026training,
      title={Training Language Models to Explain Their Own Computations}, 
      author={Belinda Z. Li and Zifan Carl Guo and Vincent Huang and Jacob Steinhardt and Jacob Andreas},
      year={2026},
      eprint={2511.08579},
      archivePrefix={arXiv},
      primaryClass={cs.CL},
      url={https://arxiv.org/abs/2511.08579}, 
}

@misc{yang2024verbalizedconfidence,
      title={On Verbalized Confidence Scores for LLMs}, 
      author={Daniel Yang and Yao-Hung Hubert Tsai and Makoto Yamada},
      year={2024},
      eprint={2412.14737},
      archivePrefix={arXiv},
      primaryClass={cs.CL},
      url={https://arxiv.org/abs/2412.14737}, 
}

@misc{kadavath2022language,
      title={Language Models (Mostly) Know What They Know}, 
      author={Saurav Kadavath and Tom Conerly and Amanda Askell and Tom Henighan and Dawn Drain and Ethan Perez and Nicholas Schiefer and Zac Hatfield-Dodds and Nova DasSarma and Eli Tran-Johnson and Scott Johnston and Sheer El-Showk and Andy Jones and Nelson Elhage and Tristan Hume and Anna Chen and Yuntao Bai and Sam Bowman and Stanislav Fort and Deep Ganguli and Danny Hernandez and Josh Jacobson and Jackson Kernion and Shauna Kravec and Liane Lovitt and Kamal Ndousse and Catherine Olsson and Sam Ringer and Dario Amodei and Tom Brown and Jack Clark and Nicholas Joseph and Ben Mann and Sam McCandlish and Chris Olah and Jared Kaplan},
      year={2022},
      eprint={2207.05221},
      archivePrefix={arXiv},
      primaryClass={cs.CL},
      url={https://arxiv.org/abs/2207.05221}, 
}

@misc{lin2022teaching,
      title={Teaching Models to Express Their Uncertainty in Words}, 
      author={Stephanie Lin and Jacob Hilton and Owain Evans},
      year={2022},
      eprint={2205.14334},
      archivePrefix={arXiv},
      primaryClass={cs.CL},
      url={https://arxiv.org/abs/2205.14334}, 
}

@misc{xiong2024llmsuncertainty,
      title={Can LLMs Express Their Uncertainty? An Empirical Evaluation of Confidence Elicitation in LLMs}, 
      author={Miao Xiong and Zhiyuan Hu and Xinyang Lu and Yifei Li and Jie Fu and Junxian He and Bryan Hooi},
      year={2024},
      eprint={2306.13063},
      archivePrefix={arXiv},
      primaryClass={cs.CL},
      url={https://arxiv.org/abs/2306.13063}, 
}

@article{Cohen1960ACO,
  title={A Coefficient of Agreement for Nominal Scales},
  author={Jacob Cohen},
  journal={Educational and Psychological Measurement},
  year={1960},
  volume={20},
  pages={37 - 46},
  url={https://api.semanticscholar.org/CorpusID:15926286}
}

@misc{aggarwal2026evaluatingchainofthoughtreasoningreusability,
      title={Evaluating Chain-of-Thought Reasoning through Reusability and Verifiability}, 
      author={Shashank Aggarwal and Ram Vikas Mishra and Amit Awekar},
      year={2026},
      eprint={2602.17544},
      archivePrefix={arXiv},
      primaryClass={cs.AI},
      url={https://arxiv.org/abs/2602.17544}, 
}

@misc{dhuliawala2023chainofverificationreduceshallucinationlarge,
      title={Chain-of-Verification Reduces Hallucination in Large Language Models}, 
      author={Shehzaad Dhuliawala and Mojtaba Komeili and Jing Xu and Roberta Raileanu and Xian Li and Asli Celikyilmaz and Jason Weston},
      year={2023},
      eprint={2309.11495},
      archivePrefix={arXiv},
      primaryClass={cs.CL},
      url={https://arxiv.org/abs/2309.11495}, 
}

@misc{stengeleskin2024lacielistenerawarefinetuningconfidence,
      title={LACIE: Listener-Aware Finetuning for Confidence Calibration in Large Language Models}, 
      author={Elias Stengel-Eskin and Peter Hase and Mohit Bansal},
      year={2024},
      eprint={2405.21028},
      archivePrefix={arXiv},
      primaryClass={cs.CL},
      url={https://arxiv.org/abs/2405.21028}, 
}

@misc{kim2026mitigatinglegibilitytaxdecoupled,
      title={Mitigating Legibility Tax with Decoupled Prover-Verifier Games}, 
      author={Yegon Kim and Juho Lee},
      year={2026},
      eprint={2602.23248},
      archivePrefix={arXiv},
      primaryClass={cs.AI},
      url={https://arxiv.org/abs/2602.23248}, 
}

@misc{kirchner2024proververifiergamesimprovelegibility,
      title={Prover-Verifier Games improve legibility of LLM outputs}, 
      author={Jan Hendrik Kirchner and Yining Chen and Harri Edwards and Jan Leike and Nat McAleese and Yuri Burda},
      year={2024},
      eprint={2407.13692},
      archivePrefix={arXiv},
      primaryClass={cs.CL},
      url={https://arxiv.org/abs/2407.13692}, 
}

@misc{hase2026counterfactualsimulationtrainingchainofthought,
      title={Counterfactual Simulation Training for Chain-of-Thought Faithfulness}, 
      author={Peter Hase and Christopher Potts},
      year={2026},
      eprint={2602.20710},
      archivePrefix={arXiv},
      primaryClass={cs.AI},
      url={https://arxiv.org/abs/2602.20710}, 
}

@misc{limpijankit2025counterfactualsimulatabilityllmexplanations,
      title={Counterfactual Simulatability of LLM Explanations for Generation Tasks}, 
      author={Marvin Limpijankit and Yanda Chen and Melanie Subbiah and Nicholas Deas and Kathleen McKeown},
      year={2025},
      eprint={2505.21740},
      archivePrefix={arXiv},
      primaryClass={cs.CL},
      url={https://arxiv.org/abs/2505.21740}, 
}

@misc{mayne2026positivecasefaithfulnessllm,
      title={A Positive Case for Faithfulness: LLM Self-Explanations Help Predict Model Behavior}, 
      author={Harry Mayne and Justin Singh Kang and Dewi Gould and Kannan Ramchandran and Adam Mahdi and Noah Y. Siegel},
      year={2026},
      eprint={2602.02639},
      archivePrefix={arXiv},
      primaryClass={cs.AI},
      url={https://arxiv.org/abs/2602.02639}, 
}

@misc{hong2026llmselfexplanationshelpusers,
      title={Do LLM Self-Explanations Help Users Predict Model Behavior? Evaluating Counterfactual Simulatability with Pragmatic Perturbations}, 
      author={Pingjun Hong and Benjamin Roth},
      year={2026},
      eprint={2601.03775},
      archivePrefix={arXiv},
      primaryClass={cs.CL},
      url={https://arxiv.org/abs/2601.03775}, 
}

@misc{chen2023modelsexplainthemselvescounterfactual,
      title={Do Models Explain Themselves? Counterfactual Simulatability of Natural Language Explanations}, 
      author={Yanda Chen and Ruiqi Zhong and Narutatsu Ri and Chen Zhao and He He and Jacob Steinhardt and Zhou Yu and Kathleen McKeown},
      year={2023},
      eprint={2307.08678},
      archivePrefix={arXiv},
      primaryClass={cs.CL},
      url={https://arxiv.org/abs/2307.08678}, 
}

@misc{hase2020evaluatingexplainableaialgorithmic,
      title={Evaluating Explainable AI: Which Algorithmic Explanations Help Users Predict Model Behavior?}, 
      author={Peter Hase and Mohit Bansal},
      year={2020},
      eprint={2005.01831},
      archivePrefix={arXiv},
      primaryClass={cs.CL},
      url={https://arxiv.org/abs/2005.01831}, 
}

@misc{ibrahim2025measuringmitigatingoverreliancenecessary,
      title={Measuring and mitigating overreliance is necessary for building human-compatible AI}, 
      author={Lujain Ibrahim and Katherine M. Collins and Sunnie S. Y. Kim and Anka Reuel and Max Lamparth and Kevin Feng and Lama Ahmad and Prajna Soni and Alia El Kattan and Merlin Stein and Siddharth Swaroop and Ilia Sucholutsky and Andrew Strait and Q. Vera Liao and Umang Bhatt},
      year={2025},
      eprint={2509.08010},
      archivePrefix={arXiv},
      primaryClass={cs.CY},
      url={https://arxiv.org/abs/2509.08010}, 
}

@inproceedings{schemmer2023appropriate,
   title={Appropriate Reliance on AI Advice: Conceptualization and the Effect of Explanations},
   url={http://dx.doi.org/10.1145/3581641.3584066},
   DOI={10.1145/3581641.3584066},
   booktitle={Proceedings of the 28th International Conference on Intelligent User Interfaces},
   publisher={ACM},
   author={Schemmer, Max and Kuehl, Niklas and Benz, Carina and Bartos, Andrea and Satzger, Gerhard},
   year={2023},
   month=mar, pages={410–422},
   collection={IUI ’23} }

@misc{bo2025relyrelyevaluatinginterventions,
      title={To Rely or Not to Rely? Evaluating Interventions for Appropriate Reliance on Large Language Models}, 
      author={Jessica Y. Bo and Sophia Wan and Ashton Anderson},
      year={2025},
      eprint={2412.15584},
      archivePrefix={arXiv},
      primaryClass={cs.HC},
      url={https://arxiv.org/abs/2412.15584}, 
}

@inproceedings{kim2025fostering,
   title={Fostering Appropriate Reliance on Large Language Models: The Role of Explanations, Sources, and Inconsistencies},
   url={http://dx.doi.org/10.1145/3706598.3714020},
   DOI={10.1145/3706598.3714020},
   booktitle={Proceedings of the 2025 CHI Conference on Human Factors in Computing Systems},
   publisher={ACM},
   author={Kim, Sunnie S. Y. and Vaughan, Jennifer Wortman and Liao, Q. Vera and Lombrozo, Tania and Russakovsky, Olga},
   year={2025},
   month=apr, pages={1–19},
   collection={CHI ’25} }

@misc{bansal2021doesexceedpartseffect,
      title={Does the Whole Exceed its Parts? The Effect of AI Explanations on Complementary Team Performance}, 
      author={Gagan Bansal and Tongshuang Wu and Joyce Zhou and Raymond Fok and Besmira Nushi and Ece Kamar and Marco Tulio Ribeiro and Daniel S. Weld},
      year={2021},
      eprint={2006.14779},
      archivePrefix={arXiv},
      primaryClass={cs.AI},
      url={https://arxiv.org/abs/2006.14779}, 
}

@article{bucinca2021trust,
   title={To Trust or to Think: Cognitive Forcing Functions Can Reduce Overreliance on AI in AI-assisted Decision-making},
   volume={5},
   ISSN={2573-0142},
   url={http://dx.doi.org/10.1145/3449287},
   DOI={10.1145/3449287},
   number={CSCW1},
   journal={Proceedings of the ACM on Human-Computer Interaction},
   publisher={Association for Computing Machinery (ACM)},
   author={Buçinca, Zana and Malaya, Maja Barbara and Gajos, Krzysztof Z.},
   year={2021},
   month=apr, pages={1–21} }

@misc{kang2025scalablebestofn,
      title={Scalable Best-of-N Selection for Large Language Models via Self-Certainty}, 
      author={Zhewei Kang and Xuandong Zhao and Dawn Song},
      year={2025},
      eprint={2502.18581},
      archivePrefix={arXiv},
      primaryClass={cs.CL},
      url={https://arxiv.org/abs/2502.18581}, 
}

@misc{shu2023rewritelm,
      title={RewriteLM: An Instruction-Tuned Large Language Model for Text Rewriting}, 
      author={Lei Shu and Liangchen Luo and Jayakumar Hoskere and Yun Zhu and Yinxiao Liu and Simon Tong and Jindong Chen and Lei Meng},
      year={2023},
      eprint={2305.15685},
      archivePrefix={arXiv},
      primaryClass={cs.CL},
      url={https://arxiv.org/abs/2305.15685}, 
}

@misc{ning2024skeletonofthought,
      title={Skeleton-of-Thought: Prompting LLMs for Efficient Parallel Generation}, 
      author={Xuefei Ning and Zinan Lin and Zixuan Zhou and Zifu Wang and Huazhong Yang and Yu Wang},
      year={2024},
      eprint={2307.15337},
      archivePrefix={arXiv},
      primaryClass={cs.CL},
      url={https://arxiv.org/abs/2307.15337}, 
}

@misc{deng2024rephraserespond,
      title={Rephrase and Respond: Let Large Language Models Ask Better Questions for Themselves}, 
      author={Yihe Deng and Weitong Zhang and Zixiang Chen and Quanquan Gu},
      year={2024},
      eprint={2311.04205},
      archivePrefix={arXiv},
      primaryClass={cs.CL},
      url={https://arxiv.org/abs/2311.04205}, 
}

@misc{madaan2023selfrefine,
      title={Self-Refine: Iterative Refinement with Self-Feedback}, 
      author={Aman Madaan and Niket Tandon and Prakhar Gupta and Skyler Hallinan and Luyu Gao and Sarah Wiegreffe and Uri Alon and Nouha Dziri and Shrimai Prabhumoye and Yiming Yang and Shashank Gupta and Bodhisattwa Prasad Majumder and Katherine Hermann and Sean Welleck and Amir Yazdanbakhsh and Peter Clark},
      year={2023},
      eprint={2303.17651},
      archivePrefix={arXiv},
      primaryClass={cs.CL},
      url={https://arxiv.org/abs/2303.17651}, 
}

@misc{stiennon2022learningsummarize,
      title={Learning to summarize from human feedback}, 
      author={Nisan Stiennon and Long Ouyang and Jeff Wu and Daniel M. Ziegler and Ryan Lowe and Chelsea Voss and Alec Radford and Dario Amodei and Paul Christiano},
      year={2022},
      eprint={2009.01325},
      archivePrefix={arXiv},
      primaryClass={cs.CL},
      url={https://arxiv.org/abs/2009.01325}, 
}

@inproceedings{taubenfeld2025confidence,
   title={Confidence Improves Self-Consistency in LLMs},
   url={http://dx.doi.org/10.18653/v1/2025.findings-acl.1030},
   DOI={10.18653/v1/2025.findings-acl.1030},
   booktitle={Findings of the Association for Computational Linguistics: ACL 2025},
   publisher={Association for Computational Linguistics},
   author={Taubenfeld, Amir and Sheffer, Tom and Ofek, Eran and Feder, Amir and Goldstein, Ariel and Gekhman, Zorik and Yona, Gal},
   year={2025},
   pages={20090–20111} }

@misc{ichihara2025bestofn,
      title={Evaluation of Best-of-N Sampling Strategies for Language Model Alignment}, 
      author={Yuki Ichihara and Yuu Jinnai and Tetsuro Morimura and Kaito Ariu and Kenshi Abe and Mitsuki Sakamoto and Eiji Uchibe},
      year={2025},
      eprint={2502.12668},
      archivePrefix={arXiv},
      primaryClass={cs.CL},
      url={https://arxiv.org/abs/2502.12668}, 
}
\bibliographystyle{COLM/colm2026_conference}

\appendix
\clearpage
\label{app:human_study_math_RR}

\newcommand{\tocentry}[2]{\hyperref[#1]{{\color{black}\S\ref*{#1}~~#2}} \dotfill \pageref{#1}}

\section*{Appendix Table of Contents}
\vspace{-2pt}
\noindent
\tocentry{app:checkpoint_full}{Full Post-Training Checkpoint Results}\\
\tocentry{app:stronger_models_full}{Full Cross-Model Comparison Results}\\
\tocentry{app:rephrase_prompts}{Stylistic Rephrase Prompts}\\
\tocentry{app:rr_prompts}{Reflect-and-Rephrase Prompts}\\
\tocentry{app:or_prompts}{Oracle Rephrase Prompts}\\
\tocentry{app:rephrase_full}{Full Rephrase Results}\\
\tocentry{app:calibrated_confidence}{Calibrated Confidence Detailed Results}\\
\tocentry{app:best_of_n}{Best-of-$n$ Selection Analysis}\\
\tocentry{app:model_names}{Model Names and API Identifiers}\\
\tocentry{app:data_preparation}{Response Preparation and Ground-Truth Grading Details}\\
\tocentry{app:human_study}{Human Study and Platform Implementation Details}

\section{Full Post-Training Checkpoint Results}
\label{app:checkpoint_full}

Tables~\ref{tab:checkpoint_tulu} and~\ref{tab:checkpoint_olmo} report accuracy, \vnorm{}, and the four per-cell scores (FP, TN, FN, TP) for each post-training checkpoint, averaged over the three LLM raters. The main paper (Figure~\ref{fig:checkpoint_vnorm}) plots only accuracy and \vnorm{}.

\begin{table}[ht]
\centering
\caption{Post-training checkpoint results for Tulu3.1-8B across four benchmarks. All values averaged over three LLM raters.}
\vspace{4pt}
\label{tab:checkpoint_tulu}
\footnotesize
\begin{tabular}{llcccccc}
\toprule
\textbf{Dataset} & \textbf{Stage} & \textbf{Acc} & \vnorm{} & \textbf{FP} & \textbf{TN} & \textbf{FN} & \textbf{TP} \\
\midrule
\multirow{4}{*}{GSM8K}
 & Base     & 0.324 & 0.737 & 0.703 & 0.897 & 0.633 & 0.716 \\
 & SFT      & 0.659 & 0.854 & 0.717 & 0.872 & 0.872 & 0.955 \\
 & DPO      & 0.822 & 0.823 & 0.602 & 0.824 & 0.889 & 0.977 \\
 & Instruct & 0.882 & 0.839 & 0.656 & 0.830 & 0.897 & 0.974 \\
\midrule
\multirow{4}{*}{MATH500}
 & Base     & 0.157 & 0.691 & 0.724 & 0.937 & 0.398 & 0.706 \\
 & SFT      & 0.306 & 0.755 & 0.626 & 0.899 & 0.626 & 0.869 \\
 & DPO      & 0.451 & 0.779 & 0.602 & 0.847 & 0.754 & 0.913 \\
 & Instruct & 0.488 & 0.782 & 0.606 & 0.871 & 0.743 & 0.909 \\
\midrule
\multirow{4}{*}{MMLU}
 & Base     & 0.415 & 0.557 & 0.298 & 0.842 & 0.312 & 0.777 \\
 & SFT      & 0.604 & 0.624 & 0.286 & 0.902 & 0.386 & 0.922 \\
 & DPO      & 0.638 & 0.649 & 0.299 & 0.838 & 0.531 & 0.927 \\
 & Instruct & 0.628 & 0.648 & 0.320 & 0.835 & 0.494 & 0.941 \\
\midrule
\multirow{4}{*}{MMLU-Pro}
 & Base     & 0.253 & 0.572 & 0.413 & 0.908 & 0.217 & 0.750 \\
 & SFT      & 0.338 & 0.656 & 0.485 & 0.915 & 0.324 & 0.901 \\
 & DPO      & 0.397 & 0.681 & 0.427 & 0.893 & 0.508 & 0.897 \\
 & Instruct & 0.384 & 0.700 & 0.514 & 0.862 & 0.532 & 0.891 \\
\bottomrule
\end{tabular}
\end{table}

\begin{table}[t]
\centering
\caption{Post-training checkpoint results for OLMo2-7B across four benchmarks. All values averaged over three LLM raters.}
\vspace{4pt}
\label{tab:checkpoint_olmo}
\footnotesize
\begin{tabular}{llcccccc}
\toprule
\textbf{Dataset} & \textbf{Stage} & \textbf{Acc} & \vnorm{} & \textbf{FP} & \textbf{TN} & \textbf{FN} & \textbf{TP} \\
\midrule
\multirow{4}{*}{GSM8K}
 & Base     & 0.308 & 0.672 & 0.442 & 0.909 & 0.569 & 0.766 \\
 & SFT      & 0.614 & 0.799 & 0.612 & 0.857 & 0.821 & 0.905 \\
 & DPO      & 0.749 & 0.782 & 0.574 & 0.808 & 0.844 & 0.904 \\
 & Instruct & 0.754 & 0.781 & 0.541 & 0.820 & 0.870 & 0.892 \\
\midrule
\multirow{4}{*}{MATH500}
 & Base     & 0.143 & 0.653 & 0.639 & 0.945 & 0.285 & 0.742 \\
 & SFT      & 0.222 & 0.719 & 0.713 & 0.924 & 0.439 & 0.800 \\
 & DPO      & 0.317 & 0.773 & 0.717 & 0.886 & 0.668 & 0.820 \\
 & Instruct & 0.329 & 0.765 & 0.710 & 0.907 & 0.622 & 0.821 \\
\midrule
\multirow{4}{*}{MMLU}
 & Base     & 0.358 & 0.601 & 0.476 & 0.908 & 0.239 & 0.779 \\
 & SFT      & 0.499 & 0.631 & 0.415 & 0.912 & 0.337 & 0.860 \\
 & DPO      & 0.562 & 0.648 & 0.402 & 0.891 & 0.412 & 0.888 \\
 & Instruct & 0.563 & 0.654 & 0.479 & 0.844 & 0.405 & 0.890 \\
\midrule
\multirow{4}{*}{MMLU-Pro}
 & Base     & 0.150 & 0.582 & 0.549 & 0.929 & 0.204 & 0.646 \\
 & SFT      & 0.259 & 0.656 & 0.580 & 0.950 & 0.287 & 0.808 \\
 & DPO      & 0.273 & 0.666 & 0.510 & 0.928 & 0.394 & 0.833 \\
 & Instruct & 0.286 & 0.678 & 0.581 & 0.919 & 0.364 & 0.848 \\
\bottomrule
\end{tabular}
\end{table}

\section{Full Cross-Model Comparison Results}
\label{app:stronger_models_full}

Tables~\ref{tab:stronger_math} and~\ref{tab:stronger_qa} report accuracy, \vnorm{}, and the four per-cell scores (FP, TN, FN, TP) for all seven models, averaged over the three LLM raters. The main paper (Figure~\ref{fig:acc_vs_vnorm}) plots only accuracy and \vnorm{}.

\begin{table}[t]
\centering
\caption{Full results for seven models on mathematical reasoning benchmarks. All values averaged over three LLM raters.}
\vspace{4pt}
\label{tab:stronger_math}
\footnotesize
\begin{tabular}{llcccccc}
\toprule
\textbf{Dataset} & \textbf{Model} & \textbf{Acc} & \vnorm{} & \textbf{FP} & \textbf{TN} & \textbf{FN} & \textbf{TP} \\
\midrule
\multirow{7}{*}{GSM8K}
 & Llama3.1-8B      & 0.779 & 0.804 & 0.595 & 0.764 & 0.893 & 0.963 \\
 & Tulu3.1-8B          & 0.882 & 0.788 & 0.506 & 0.768 & 0.898 & 0.982 \\
 & OLMo2-7B            & 0.754 & 0.760 & 0.469 & 0.723 & 0.886 & 0.962 \\
 & Qwen3            & 0.978 & 0.653 & 0.157 & 0.355 & 0.927 & 0.967 \\
 & DeepSeek-V3      & 0.959 & 0.569 & 0.073 & 0.263 & 0.950 & 0.991 \\
 & Llama4-Maverick  & 0.972 & 0.640 & 0.223 & 0.429 & 0.916 & 0.991 \\
 & Grok-4           & 0.927 & 0.627 & 0.071 & 0.500 & 0.942 & 0.997 \\
\midrule
\multirow{7}{*}{MATH500}
 & Llama3.1-8B      & 0.465 & 0.764 & 0.618 & 0.803 & 0.756 & 0.877 \\
 & Tulu3.1-8B          & 0.488 & 0.782 & 0.606 & 0.871 & 0.743 & 0.909 \\
 & OLMo2-7B            & 0.329 & 0.765 & 0.710 & 0.907 & 0.622 & 0.821 \\
 & Qwen3            & 0.980 & 0.516 & 0.000 & 0.175 & 0.921 & 0.969 \\
 & DeepSeek-V3      & 0.985 & 0.567 & 0.014 & 0.287 & 0.971 & 0.997 \\
 & Llama4-Maverick  & 0.903 & 0.833 & 0.779 & 0.812 & 0.803 & 0.936 \\
 & Grok-4           & 0.787 & 0.765 & 0.515 & 0.733 & 0.854 & 0.957 \\
\bottomrule
\end{tabular}
\end{table}

\begin{table}[t]
\centering
\caption{Full results for seven models on factual knowledge QA benchmarks. All values averaged over three LLM raters.}
\vspace{4pt}
\label{tab:stronger_qa}
\footnotesize
\begin{tabular}{llcccccc}
\toprule
\textbf{Dataset} & \textbf{Model} & \textbf{Acc} & \vnorm{} & \textbf{FP} & \textbf{TN} & \textbf{FN} & \textbf{TP} \\
\midrule
\multirow{7}{*}{MMLU}
 & Llama3.1-8B      & 0.624 & 0.629 & 0.329 & 0.657 & 0.583 & 0.949 \\
 & Tulu3.1-8B          & 0.628 & 0.648 & 0.320 & 0.835 & 0.494 & 0.941 \\
 & OLMo2-7B            & 0.563 & 0.654 & 0.479 & 0.844 & 0.405 & 0.890 \\
 & Qwen3            & 0.712 & 0.639 & 0.426 & 0.568 & 0.635 & 0.925 \\
 & DeepSeek-V3      & 0.833 & 0.563 & 0.052 & 0.275 & 0.932 & 0.995 \\
 & Llama4-Maverick  & 0.771 & 0.590 & 0.166 & 0.353 & 0.853 & 0.987 \\
 & Grok-4           & 0.716 & 0.539 & 0.073 & 0.198 & 0.899 & 0.988 \\
\midrule
\multirow{7}{*}{MMLU-Pro}
 & Llama3.1-8B      & 0.404 & 0.626 & 0.382 & 0.706 & 0.505 & 0.912 \\
 & Tulu3.1-8B          & 0.384 & 0.700 & 0.514 & 0.862 & 0.532 & 0.891 \\
 & OLMo2-7B            & 0.286 & 0.678 & 0.581 & 0.919 & 0.364 & 0.848 \\
 & Qwen3            & 0.506 & 0.663 & 0.359 & 0.662 & 0.711 & 0.921 \\
 & DeepSeek-V3      & 0.794 & 0.620 & 0.175 & 0.394 & 0.923 & 0.988 \\
 & Llama4-Maverick  & 0.800 & 0.639 & 0.304 & 0.440 & 0.841 & 0.971 \\
 & Grok-4           & 0.694 & 0.615 & 0.166 & 0.423 & 0.889 & 0.983 \\
\midrule
\multirow{7}{*}{TruthfulQA}
 & Llama3.1-8B      & 0.553 & 0.640 & 0.295 & 0.720 & 0.606 & 0.940 \\
 & Tulu3.1-8B          & 0.390 & 0.667 & 0.321 & 0.923 & 0.473 & 0.951 \\
 & OLMo2-7B            & 0.418 & 0.669 & 0.399 & 0.833 & 0.512 & 0.934 \\
 & Qwen3            & 0.660 & 0.673 & 0.602 & 0.792 & 0.448 & 0.851 \\
 & DeepSeek-V3      & 0.833 & 0.572 & 0.103 & 0.400 & 0.811 & 0.977 \\
 & Llama4-Maverick  & 0.724 & 0.603 & 0.179 & 0.548 & 0.725 & 0.961 \\
 & Grok-4           & 0.777 & 0.553 & 0.074 & 0.277 & 0.876 & 0.987 \\
\bottomrule
\end{tabular}
\end{table}

\section{Stylistic Rephrase Prompts}
\label{app:rephrase_prompts}

Below are the full prompts used for each stylistic rephrase method. In each case, \texttt{\{question\}} and \texttt{\{response\}} are replaced with the original question and model response.

\paragraph{Structured.}
\begin{quote}\ttfamily\small
Original Question: \{question\}

Original Response: \{response\}

Task: Rewrite the response as a sequence of clearly numbered steps so that each step can be read and verified independently. Specifically:\\
- Break the reasoning into numbered steps (Step 1, Step 2, ...)\\
- Each step should contain exactly one atomic reasoning action: one calculation, one factual claim, or one logical deduction\\
- Make implicit steps explicit --- if the original skips an intermediate calculation or assumption, add it as its own numbered step\\
- Each step should be one to two sentences and self-contained\\
- Do NOT add new reasoning, change any values, or alter the conclusion\\
Important: Keep the final answer exactly the same (after `\#\#\#\#' if present), placed after the last numbered step.

Rephrased Response:
\end{quote}

\paragraph{Professional.}
\begin{quote}\ttfamily\small
Original Question: \{question\}

Original Response: \{response\}

Task: Rewrite the response in a professional, precise style while preserving the exact same reasoning steps and final answer. Specifically:\\
- Use consistent terminology throughout (do not alternate between synonyms for the same concept)\\
- Add clear logical connectives to link reasoning steps (e.g., `therefore', `it follows that', `consequently', `since')\\
- Use formal, neutral language --- avoid colloquial phrasing, filler words, and informal constructions\\
- Ensure each sentence is precise and unambiguous\\
- Do NOT add new reasoning steps, change any values, or alter the conclusion\\
Important: Keep the final answer exactly the same (after `\#\#\#\#' if present).

Rephrased Response:
\end{quote}

\paragraph{Simplified.}
\begin{quote}\ttfamily\small
Original Question: \{question\}

Original Response: \{response\}

Task: Rewrite the response to be more concise by removing redundancy, while preserving every distinct reasoning step and the final answer. Specifically, remove:\\
- Restatements of the question or prior steps\\
- Filler phrases (e.g., `It is important to note that', `As we can see', `Therefore, we can conclude that')\\
- Repetition of values or facts already stated earlier\\
- Transitional padding that does not add logical content\\
Keep:\\
- Every distinct reasoning step, even if expressed more briefly\\
- All numerical values and intermediate calculations\\
- The logical structure and order of the original reasoning\\
Do NOT remove any step that contributes to reaching the final answer, and do NOT change any values or the conclusion.\\
Important: Keep the final answer exactly the same (after `\#\#\#\#' if present).

Rephrased Response:
\end{quote}

\paragraph{Uncertain (used for calibrated confidence).}
\label{app:uncertain_prompt}
\begin{quote}\ttfamily\small
Original Question: \{question\}

Original Response: \{response\}

Task: Rewrite the response to express genuine uncertainty. Convey the same information and conclusion, but acknowledge uncertainty where reasonable. Preserve any final answer markers (e.g., `\#\#\#\#') and their content.

Rephrased Response:
\end{quote}

\section{Reflect-and-Rephrase Prompts}
\label{app:rr_prompts}

The reflect-and-rephrase (\textsc{RR}) method is a two-round pipeline. In the first round (\textit{Reflect}), the rephrase model receives the target response alongside $k$ alternative responses and produces an analysis of agreements and discrepancies. In the second round (\textit{Rephrase}), the model rewrites the justification conditioned on this analysis. In both prompts, \texttt{\{question\}}, \texttt{\{response\}}, and \texttt{\{alternatives\}} are replaced with the actual inputs.

\paragraph{Reflect prompt.}
\begin{quote}\ttfamily\small
You are a helpful assistant that helps a reader judge whether the main response's answer is correct.

Your role is to analyze responses and surface potential errors, uncertainties, and points a reader should verify.

Question: \{question\}

Main Response: \{response\}

Alternative Responses: \{alternatives\}

Analyze the main response, using the alternatives as evidence:\\
1. Do the final answers agree or disagree?\\
2. What are the key steps or assumptions in the main response? Are they different from the alternatives? Could any of them be wrong?\\
3. If there are differences between responses, which reasoning seems more reliable?\\
4. How confident should we be in the main response? Is it likely correct, uncertain, or likely wrong? If uncertain or likely wrong, identify the specific problematic step(s) and what the correct steps might be.

Analysis:
\end{quote}

\paragraph{Rephrase prompt.}
\begin{quote}\ttfamily\small
You are a helpful assistant that rewrites responses to help readers judge whether the response's answer is correct. Your goal is to make potential errors and uncertainties visible, not to persuade the reader that the answer is right.

Question: \{question\}

Original Response: \{response\}

Analysis: \{analysis\}

Rewrite the original response to help readers judge whether the answer is correct.

The rewrite should NOT be a persuasive proof. Instead, it should:\\
- Explain the reasoning transparently\\
- Surface potential failure points identified in the analysis\\
- If the analysis found likely errors, note them in the steps\\
- If any steps, facts, or assumptions are uncertain, express them tentatively rather than assertively

Rules:\\
- Keep the same final answer (preserve any \#\#\#\# markers exactly)\\
- Write as a self-contained explanation---don't mention the analysis or alternatives

Rephrased Response:
\end{quote}

\section{Oracle Rephrase Prompts}
\label{app:or_prompts}

The oracle rephrase (\textsc{OR}) method is a three-step pipeline. In Step~1, the rephrase model (Tulu3.1-8B) extracts all verifiable claims from the justification. In Step~2, an oracle model (Claude-Sonnet-4.5) independently judges each claim as \textsc{Correct}, \textsc{Incorrect}, or \textsc{Not\_Verifiable}. In Step~3, the rephrase model rewrites the justification with explicit inline annotations at flagged claims. Below are the prompts for each step; \texttt{\{question\}} and \texttt{\{response\}} are replaced with the actual inputs.

\paragraph{Step 1: Claim extraction (rephrase model).}
\begin{quote}\ttfamily\small
You are a helpful assistant that extracts claims from a response for fact-checking.

QUESTION: \{question\}

RESPONSE: \{response\}

Extract all the claims made in the response that can potentially be verified or checked.
Focus on:\\
- Factual statements (numbers, dates, definitions, properties)\\
- Reasoning steps and logical inferences\\
- Mathematical calculations and their results\\
- Causal claims (X causes Y, X leads to Y)\\
- References to concepts, rules, or principles

Format your output using EXACTLY this format, with each claim on its own line:\\
CLAIM \#1: [First claim text here]\\
CLAIM \#2: [Second claim text here]\\
...

Each claim should be self-contained and understandable without the full context.

Claims:
\end{quote}

\paragraph{Step 2: Claim verification (oracle model).}
This prompt is issued once per extracted claim. \texttt{\{claim\}} and \texttt{\{claim\_index\}} are replaced with the claim text and its index.
\begin{quote}\ttfamily\small
You are an expert fact-checker and reasoning verifier.

ORIGINAL QUESTION: \{question\}

CLAIM TO VERIFY (Claim \#\{claim\_index\}): \{claim\}

CONTEXT (the full response this claim comes from): \{response\}

Your task is to judge whether this claim is correct, incorrect, or not verifiable.

Provide your judgment in the following format:\\
JUDGMENT: [CORRECT / INCORRECT / NOT\_VERIFIABLE]\\
EXPLANATION: [Brief explanation of why you made this judgment]

Guidelines:\\
- CORRECT: The claim is factually accurate and logically sound.\\
- INCORRECT: The claim contains a factual error, calculation mistake, or logical flaw.\\
- NOT\_VERIFIABLE: The claim cannot be verified without additional context or external knowledge, or it is a matter of interpretation/opinion.

Be concise but precise in your explanation. Focus on the specific issue if incorrect.

Your response:
\end{quote}

\paragraph{Step 3: Rephrase with oracle notes (rephrase model).}
\texttt{\{oracle\_notes\}} is replaced with the formatted claim--judgment pairs from Step~2.
\begin{quote}\ttfamily\small
You are a helpful assistant that rewrites responses to help readers judge correctness.
An oracle verifier has analyzed the claims in the response. Your task is to incorporate this verification information into a clear, readable rewrite.

QUESTION: \{question\}

ORIGINAL RESPONSE: \{response\}

ORACLE VERIFICATION RESULTS: \{oracle\_notes\}

Rewrite the ORIGINAL RESPONSE with oracle verification notes EXPLICITLY shown.

Guidelines:\\
- Keep the original response's structure and final answer.\\
- For claims marked CORRECT: State them as-is, no annotation needed.\\
- For claims marked INCORRECT: Add an explicit note in parentheses or brackets right after the claim, showing the issue. Format: (NOTE: [issue description])\\
- For claims marked NOT\_VERIFIABLE: Add an explicit note in parentheses or brackets. Format: (NOTE: This claim could not be verified)\\
- The oracle notes should be VISIBLE and EXPLICIT---do NOT fold them naturally into the text.\\
- Preserve the final answer marker (e.g., \#\#\#\#) exactly.

Rephrased Response:
\end{quote}

\section{Full Rephrase Results}
\label{app:rephrase_full}

Tables~\ref{tab:rephrase_math_full} and~\ref{tab:rephrase_science_full} report the complete per-cell \vnorm{} breakdowns (FP, TN, FN, TP) for all rephrase methods. The main paper (Table~\ref{tab:rephrase}) reports only \vnorm{} for compactness.

\begin{table}[t]
\centering
\caption{Full per-cell results for rephrase methods on mathematical benchmarks ($\Delta$ relative to Base).}
\vspace{4pt}
\label{tab:rephrase_math_full}
\footnotesize
\begin{tabular}{llccccc}
\toprule
\textbf{Model} & & \vnorm{} & \textbf{FP} & \textbf{TN} & \textbf{FN} & \textbf{TP} \\
\midrule
\multicolumn{7}{c}{\textit{MATH500}} \\
\midrule
\multirow{5}{*}{Llama3.1-8B}
 & Base                           & 0.763    & 0.652    & 0.762    & 0.729    & 0.907 \\
 & \textsc{Prof.} $\Delta$        & $+$0.028 & $+$0.057 & $+$0.077 & $-$0.019 & $-$0.004 \\
 & \textsc{Struct.} $\Delta$      & $+$0.012 & $+$0.034 & $+$0.090 & $-$0.033 & $-$0.041 \\
 & \textsc{Simpl.} $\Delta$       & $+$0.002 & $+$0.031 & $+$0.082 & $-$0.072 & $-$0.036 \\
 & \textsc{RR} $\Delta$           & $+$0.040 & $+$0.114 & $+$0.113 & $-$0.050 & $-$0.019 \\
\midrule
\multirow{5}{*}{Tulu3.1-8B}
 & Base                           & 0.778    & 0.642    & 0.820    & 0.741    & 0.911 \\
 & \textsc{Prof.} $\Delta$        & $-$0.003 & $-$0.033 & $-$0.007 & $+$0.027 & $+$0.002 \\
 & \textsc{Struct.} $\Delta$      & $-$0.010 & $+$0.029 & $+$0.038 & $-$0.076 & $-$0.032 \\
 & \textsc{Simpl.} $\Delta$       & $-$0.017 & $-$0.001 & $+$0.027 & $-$0.074 & $-$0.019 \\
 & \textsc{RR} $\Delta$           & $+$0.024 & $+$0.066 & $+$0.046 & $-$0.011 & $-$0.005 \\
\midrule
\multirow{5}{*}{OLMo2-7B}
 & Base                           & 0.734    & 0.677    & 0.883    & 0.549    & 0.826 \\
 & \textsc{Prof.} $\Delta$        & $+$0.026 & $-$0.035 & $-$0.000 & $+$0.083 & $+$0.056 \\
 & \textsc{Struct.} $\Delta$      & $+$0.030 & $+$0.039 & $+$0.017 & $+$0.050 & $+$0.013 \\
 & \textsc{Simpl.} $\Delta$       & $+$0.020 & $-$0.015 & $+$0.009 & $+$0.052 & $+$0.033 \\
 & \textsc{RR} $\Delta$           & $+$0.061 & $+$0.056 & $+$0.007 & $+$0.107 & $+$0.073 \\
\midrule
\multicolumn{7}{c}{\textit{GSM8K}} \\
\midrule
\multirow{5}{*}{Llama3.1-8B}
 & Base                           & 0.773    & 0.562    & 0.781    & 0.809    & 0.938 \\
 & \textsc{Prof.} $\Delta$        & $-$0.040 & $-$0.186 & $+$0.034 & $-$0.017 & $+$0.007 \\
 & \textsc{Struct.} $\Delta$      & $-$0.047 & $-$0.179 & $+$0.060 & $-$0.067 & $-$0.001 \\
 & \textsc{Simpl.} $\Delta$       & $-$0.080 & $-$0.190 & $+$0.024 & $-$0.141 & $-$0.015 \\
 & \textsc{RR} $\Delta$           & $+$0.027 & $+$0.071 & $+$0.024 & $+$0.001 & $+$0.012 \\
\midrule
\multirow{5}{*}{Tulu3.1-8B}
 & Base                           & 0.757    & 0.424    & 0.827    & 0.827    & 0.951 \\
 & \textsc{Prof.} $\Delta$        & $-$0.023 & $-$0.075 & $-$0.008 & $-$0.014 & $+$0.006 \\
 & \textsc{Struct.} $\Delta$      & $-$0.030 & $-$0.061 & $-$0.005 & $-$0.048 & $-$0.008 \\
 & \textsc{Simpl.} $\Delta$       & $-$0.059 & $-$0.099 & $-$0.006 & $-$0.106 & $-$0.026 \\
 & \textsc{RR} $\Delta$           & $+$0.048 & $+$0.182 & $-$0.031 & $+$0.033 & $+$0.008 \\
\midrule
\multirow{5}{*}{OLMo2-7B}
 & Base                           & 0.736    & 0.521    & 0.774    & 0.762    & 0.889 \\
 & \textsc{Prof.} $\Delta$        & $-$0.015 & $-$0.082 & $+$0.003 & $-$0.001 & $+$0.020 \\
 & \textsc{Struct.} $\Delta$      & $-$0.017 & $-$0.114 & $+$0.023 & $+$0.013 & $+$0.008 \\
 & \textsc{Simpl.} $\Delta$       & $-$0.058 & $-$0.174 & $-$0.010 & $-$0.059 & $+$0.009 \\
 & \textsc{RR} $\Delta$           & $+$0.083 & $+$0.168 & $+$0.025 & $+$0.087 & $+$0.054 \\
\bottomrule
\end{tabular}
\end{table}

\begin{table}[t]
\centering
\caption{Full per-cell results for rephrase methods on factual knowledge QA benchmarks ($\Delta$ relative to Base).}
\vspace{4pt}
\label{tab:rephrase_science_full}
\footnotesize
\begin{tabular}{llccccc}
\toprule
\textbf{Model} & & \vnorm{} & \textbf{FP} & \textbf{TN} & \textbf{FN} & \textbf{TP} \\
\midrule
\multicolumn{7}{c}{\textit{MMLU}} \\
\midrule
\multirow{6}{*}{Llama3.1-8B}
 & Base                           & 0.625    & 0.293    & 0.679    & 0.585    & 0.944 \\
 & \textsc{Prof.} $\Delta$        & $+$0.029 & $+$0.058 & $+$0.092 & $-$0.031 & $-$0.004 \\
 & \textsc{Struct.} $\Delta$      & $+$0.037 & $+$0.072 & $+$0.062 & $+$0.013 & $+$0.001 \\
 & \textsc{Simpl.} $\Delta$       & $+$0.043 & $+$0.090 & $+$0.115 & $-$0.031 & $-$0.002 \\
 & \textsc{RR} $\Delta$           & $+$0.036 & $+$0.069 & $+$0.049 & $+$0.004 & $+$0.021 \\
 & \textsc{OR} $\Delta$           & $+$0.086 & $+$0.216 & $+$0.139 & $-$0.023 & $+$0.014 \\
\midrule
\multirow{6}{*}{Tulu3.1-8B}
 & Base                           & 0.657    & 0.397    & 0.796    & 0.517    & 0.918 \\
 & \textsc{Prof.} $\Delta$        & $+$0.009 & $+$0.006 & $+$0.030 & $-$0.003 & $+$0.000 \\
 & \textsc{Struct.} $\Delta$      & $+$0.002 & $-$0.051 & $-$0.032 & $+$0.085 & $+$0.007 \\
 & \textsc{Simpl.} $\Delta$       & $+$0.000 & $-$0.013 & $+$0.015 & $-$0.005 & $+$0.004 \\
 & \textsc{RR} $\Delta$           & $+$0.000 & $+$0.023 & $-$0.076 & $+$0.035 & $+$0.020 \\
 & \textsc{OR} $\Delta$           & $+$0.066 & $+$0.191 & $+$0.029 & $-$0.001 & $+$0.046 \\
\midrule
\multirow{6}{*}{OLMo2-7B}
 & Base                           & 0.640    & 0.483    & 0.835    & 0.370    & 0.872 \\
 & \textsc{Prof.} $\Delta$        & $+$0.019 & $-$0.008 & $-$0.003 & $+$0.067 & $+$0.020 \\
 & \textsc{Struct.} $\Delta$      & $+$0.013 & $-$0.058 & $-$0.074 & $+$0.140 & $+$0.042 \\
 & \textsc{Simpl.} $\Delta$       & $+$0.002 & $-$0.026 & $-$0.010 & $+$0.026 & $+$0.019 \\
 & \textsc{RR} $\Delta$           & $+$0.006 & $-$0.072 & $-$0.129 & $+$0.168 & $+$0.059 \\
 & \textsc{OR} $\Delta$           & $+$0.045 & $+$0.094 & $-$0.033 & $+$0.083 & $+$0.038 \\
\midrule
\multicolumn{7}{c}{\textit{TruthfulQA}} \\
\midrule
\multirow{6}{*}{Llama3.1-8B}
 & Base                           & 0.636    & 0.311    & 0.707    & 0.612    & 0.914 \\
 & \textsc{Prof.} $\Delta$        & $+$0.009 & $-$0.024 & $+$0.087 & $-$0.036 & $+$0.011 \\
 & \textsc{Struct.} $\Delta$      & $+$0.022 & $-$0.010 & $+$0.055 & $+$0.021 & $+$0.020 \\
 & \textsc{Simpl.} $\Delta$       & $+$0.026 & $+$0.046 & $+$0.087 & $-$0.034 & $+$0.005 \\
 & \textsc{RR} $\Delta$           & $+$0.007 & $-$0.101 & $-$0.016 & $+$0.100 & $+$0.043 \\
 & \textsc{OR} $\Delta$           & $+$0.069 & $+$0.075 & $+$0.145 & $+$0.014 & $+$0.043 \\
\midrule
\multirow{6}{*}{Tulu3.1-8B}
 & Base                           & 0.685    & 0.420    & 0.892    & 0.510    & 0.917 \\
 & \textsc{Prof.} $\Delta$        & $-$0.004 & $-$0.037 & $-$0.003 & $+$0.014 & $+$0.009 \\
 & \textsc{Struct.} $\Delta$      & $+$0.013 & $-$0.062 & $-$0.101 & $+$0.172 & $+$0.044 \\
 & \textsc{Simpl.} $\Delta$       & $+$0.002 & $-$0.026 & $-$0.022 & $+$0.039 & $+$0.015 \\
 & \textsc{RR} $\Delta$           & $-$0.069 & $-$0.225 & $-$0.105 & $+$0.052 & $+$0.002 \\
 & \textsc{OR} $\Delta$           & $+$0.038 & $+$0.048 & $+$0.028 & $+$0.037 & $+$0.039 \\
\midrule
\multirow{6}{*}{OLMo2-7B}
 & Base                           & 0.670    & 0.446    & 0.819    & 0.531    & 0.884 \\
 & \textsc{Prof.} $\Delta$        & $-$0.011 & $-$0.024 & $+$0.023 & $-$0.042 & $-$0.002 \\
 & \textsc{Struct.} $\Delta$      & $-$0.014 & $-$0.071 & $+$0.000 & $+$0.005 & $+$0.010 \\
 & \textsc{Simpl.} $\Delta$       & $-$0.011 & $-$0.010 & $-$0.000 & $-$0.026 & $-$0.007 \\
 & \textsc{RR} $\Delta$           & $-$0.027 & $-$0.167 & $-$0.082 & $+$0.096 & $+$0.047 \\
 & \textsc{OR} $\Delta$           & $+$0.026 & $-$0.001 & $+$0.046 & $+$0.013 & $+$0.045 \\
\bottomrule
\end{tabular}
\end{table}

\section{Calibrated Confidence Detailed Results}
\label{app:calibrated_confidence}

\begin{figure}[t]
\centering
\includegraphics[width=\textwidth]{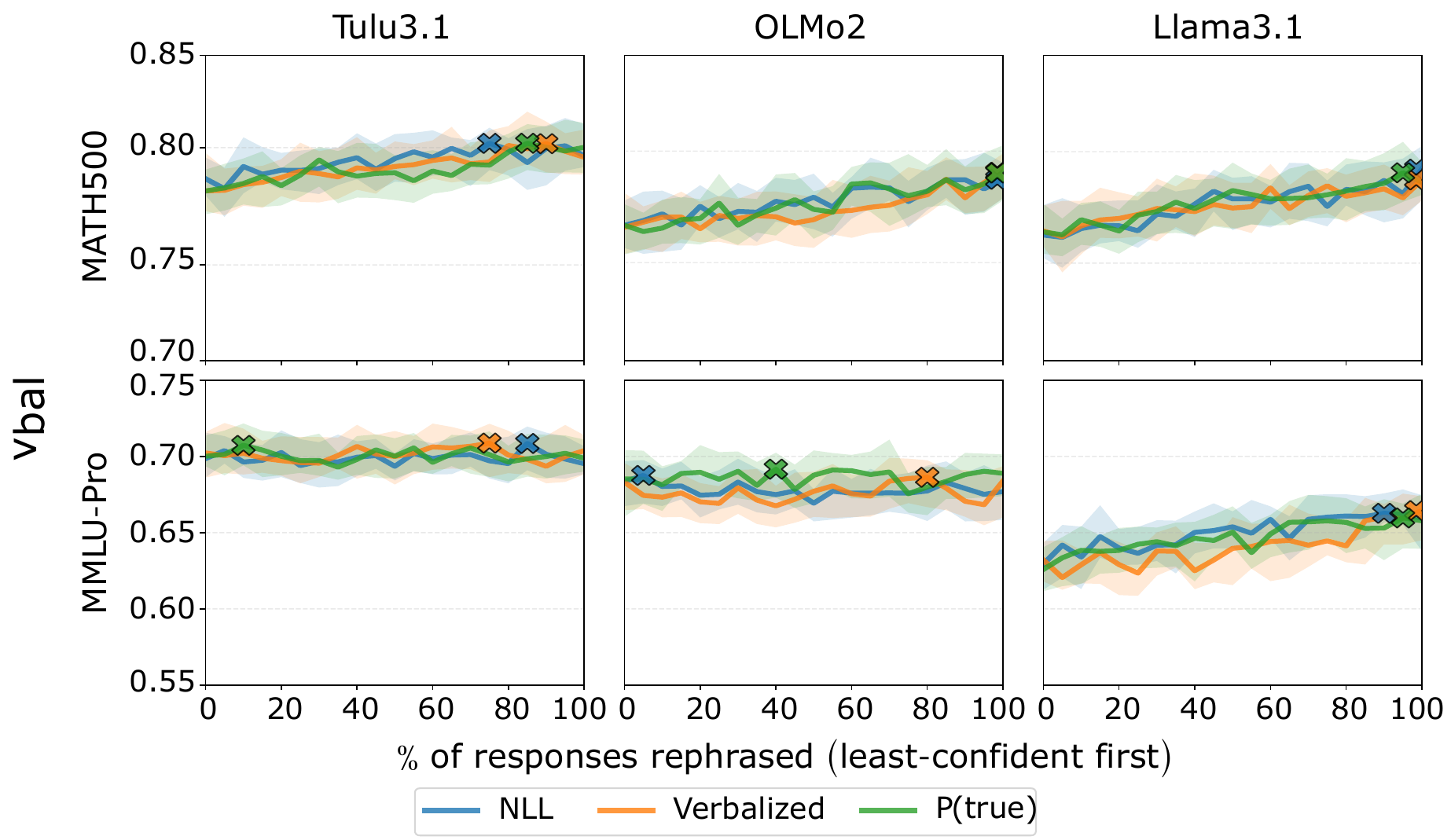}
\caption{Effect of calibrating linguistic confidence on \vnorm{} across three models and benchmarks (MATH500, MMLU, MMLU-Pro). Each curve sweeps the fraction $k$ of least-confident responses that were rephrased to express uncertainty; {\textcolor{confNLL}{$\boldsymbol{\times}$}},~{\textcolor{confVerb}{$\boldsymbol{\times}$}},~{\textcolor{confPtrue}{$\boldsymbol{\times}$}} mark the $k$ maximizing \vnorm{} for each confidence method. Optima predominantly cluster around $k{=}100\%$, suggesting that uniform hedging is as effective as targeted uncertainty calibration.}
\label{fig:calibrated_confidence}
\end{figure}

\section{Best-of-$n$ Selection Analysis}
\label{app:best_of_n}

Best-of-$n$ selection is a common inference-time strategy in which a model generates $n$ candidate responses and a scoring function selects the best one~\citep{stiennon2022learningsummarize, kang2025scalablebestofn, taubenfeld2025confidence, ichihara2025bestofn}. We test whether best-of-$n$ selection can improve verifiability.
For each question, we generate 20 candidate responses and randomly sample $n{=}5$; a selection strategy then picks one.
We evaluate on MATH500 and MMLU-Pro with three response models (Llama3.1-8B, Tulu3.1-8B, OLMo2-7B), using Gemini-2.5-Flash-Lite as the LLM rater.
Each configuration is repeated across 10 random experiments of 200 questions; Table~\ref{tab:best_of_n} reports mean accuracy and \vnorm{}.

We test nine selection strategies:
\textsc{Random} uniformly samples one candidate;
\textsc{Min/Max Len.} selects the shortest or longest response;
\textsc{Min/Max Steps} selects the response with fewest or most reasoning steps;
\textsc{Best NLL} selects the response with lowest negative log-likelihood under the response model;
\textsc{Best P(True)} selects the response with the highest model-estimated probability of correctness;
\textsc{Best Verb.\ Conf.} selects the response with the highest verbalized confidence;
and \textsc{Model Sel.} prompts the response model to directly choose the best response.

\begin{table}[t]
\centering
\caption{Best-of-5 selection results on MATH500 and MMLU-Pro ($\Delta$ vs.\ \textsc{Random}). Values are means over 10 experiments of 200 questions each. \textsc{Model Sel.} was not evaluated on MMLU-Pro.}
\vspace{4pt}
\label{tab:best_of_n}
\footnotesize
\setlength{\tabcolsep}{3.5pt}
\begin{tabular}{@{}lrrrrrr@{}}
\toprule
& \multicolumn{2}{c}{\textbf{Llama3.1-8B}} & \multicolumn{2}{c}{\textbf{Tulu3.1-8B}} & \multicolumn{2}{c}{\textbf{OLMo2-7B}} \\
\cmidrule(lr){2-3} \cmidrule(lr){4-5} \cmidrule(lr){6-7}
\textbf{Selection} & Acc & \vnorm{} & Acc & \vnorm{} & Acc & \vnorm{} \\
\midrule
\multicolumn{7}{c}{\textit{MATH500}} \\
\midrule
\textsc{Random}              & 0.430 & 0.762 & 0.446 & 0.797 & 0.310 & 0.717 \\
\textsc{Min Len.} $\Delta$            & $-$0.004 & $+$0.010 & $-$0.012 & $-$0.104 & $+$0.020 & $+$0.051 \\
\textsc{Max Len.} $\Delta$            & $-$0.022 & $+$0.001 & $+$0.020 & $+$0.043 & $+$0.004 & $+$0.012 \\
\textsc{Min Steps} $\Delta$           & $-$0.010 & $-$0.029 & $-$0.010 & $-$0.043 & $+$0.004 & $+$0.070 \\
\textsc{Max Steps} $\Delta$           & $-$0.010 & $+$0.026 & $+$0.040 & $+$0.037 & $+$0.038 & $+$0.047 \\
\textsc{Best NLL} $\Delta$            & $+$0.000 & $+$0.000 & $+$0.000 & $+$0.000 & $+$0.000 & $+$0.000 \\
\textsc{Best P(True)} $\Delta$        & $+$0.040 & $-$0.050 & $+$0.088 & $+$0.006 & $+$0.040 & $+$0.010 \\
\textsc{Best Verb.\ Conf.} $\Delta$   & $-$0.032 & $-$0.003 & $+$0.046 & $+$0.026 & $+$0.006 & $+$0.003 \\
\textsc{Model Sel.} $\Delta$          & $+$0.041 & $-$0.014 & $+$0.018 & $+$0.019 & $+$0.008 & $+$0.005 \\
\midrule
\multicolumn{7}{c}{\textit{MMLU-Pro}} \\
\midrule
\textsc{Random}              & 0.388 & 0.623 & 0.382 & 0.681 & 0.247 & 0.648 \\
\textsc{Min Len.} $\Delta$            & $+$0.043 & $+$0.013 & $-$0.007 & $+$0.037 & $+$0.008 & $+$0.010 \\
\textsc{Max Len.} $\Delta$            & $-$0.040 & $-$0.024 & $-$0.021 & $-$0.036 & $+$0.032 & $+$0.013 \\
\textsc{Min Steps} $\Delta$           & $+$0.033 & $+$0.006 & $-$0.006 & $+$0.044 & $+$0.004 & $-$0.016 \\
\textsc{Max Steps} $\Delta$           & $-$0.001 & $-$0.015 & $-$0.012 & $-$0.008 & $+$0.016 & $+$0.027 \\
\textsc{Best NLL} $\Delta$            & $+$0.051 & $-$0.011 & $+$0.025 & $+$0.020 & $+$0.021 & $+$0.008 \\
\textsc{Best P(True)} $\Delta$        & $+$0.030 & $-$0.029 & $+$0.021 & $+$0.002 & $+$0.062 & $+$0.060 \\
\textsc{Best Verb.\ Conf.} $\Delta$   & $+$0.007 & $-$0.014 & $+$0.040 & $+$0.013 & $+$0.011 & $+$0.017 \\
\textsc{Model Sel.} $\Delta$          & ---   & ---   & ---   & ---   & ---   & ---   \\
\bottomrule
\end{tabular}
\end{table}

No selection strategy consistently improves \vnorm{} across all models and datasets (Table~\ref{tab:best_of_n}).
Although \textsc{Best P(True)} raises the MATH500 accuracy of Tulu3.1-8B by $+$0.088, the corresponding \vnorm{} barely changes ($+$0.006).
In several cases, higher accuracy comes at the cost of lower \vnorm{}: on MATH500, \textsc{Best P(True)} improves the accuracy of Llama3.1-8B by $+$0.040 while reducing \vnorm{} by $-$0.050.
Similarly, \textsc{Min Len.} degrades \vnorm{} for Tulu3.1-8B on MATH500 by $-$0.104 while improving it for OLMo2-7B by $+$0.051.
Overall, these results support our finding that accuracy and verifiability are distinct dimensions of response quality, and that selecting for one does not reliably improve the other.

\section{Model Names and API Identifiers}
\label{app:model_names}

Table~\ref{tab:model_names} lists all models used in this work along with their API identifiers and roles.

\begin{table}[h]
\centering
\caption{Mapping between model names used in this paper and their API identifiers.}
\vspace{4pt}
\label{tab:model_names}
\footnotesize
\begin{tabular}{lll}
\toprule
\textbf{Paper Name} & \textbf{API Identifier} & \textbf{Role} \\
\midrule
GPT-4.1-mini & \href{https://developers.openai.com/api/docs/models/gpt-4.1-mini}{\texttt{gpt-4.1-mini-2025-04-14}} & Rater \\
Claude-Haiku-4.5 & \href{https://www.anthropic.com/claude/haiku}{\texttt{claude-haiku-4-5-20251001}} & Rater \\
Gemini-2.5-Flash-Lite & \href{https://ai.google.dev/gemini-api/docs/models/gemini-2.5-flash-lite}{\texttt{gemini-2.5-flash-lite}} & Rater \\
\midrule
GPT-5.2 & \href{https://openai.com/index/introducing-gpt-5-2/}{\texttt{gpt-5.2-2025-12-11}} & Grading \\
Claude-Sonnet-4.5 & \href{https://www.anthropic.com/news/claude-sonnet-4-5}{\texttt{claude-sonnet-4-5}} & Fact checker \\
\midrule
Llama3.1-8B & \href{https://huggingface.co/meta-llama/Llama-3.1-8B}{\texttt{meta-llama/Llama-3.1-8B-Instruct}} & Response model \\
Tulu3.1-8B & \href{https://huggingface.co/allenai/Llama-3.1-Tulu-3.1-8B}{\texttt{allenai/Llama-3.1-Tulu-3.1-8B}} & Response / Rephrase model \\
OLMo2-7B & \href{https://huggingface.co/allenai/OLMo-2-1124-7B}{\texttt{allenai/OLMo-2-1124-7B-Instruct}} & Response model \\
Qwen3 & \href{https://huggingface.co/Qwen/Qwen3-8B}{\texttt{Qwen/Qwen3-8B}} & Response model \\
DeepSeek-V3 & \href{https://github.com/deepseek-ai/deepseek-v3}{\texttt{deepseek-chat}} & Response model \\
Llama4-Maverick & \href{https://huggingface.co/meta-llama/Llama-4-Maverick-17B-128E-Instruct}{\texttt{meta-llama/Llama-4-Maverick-17B-128E-Instruct}} & Response model \\
Grok-4 & \href{https://x.ai/news/grok-4}{\texttt{grok-4}} & Response model \\
\bottomrule
\end{tabular}
\end{table}

\section{Response Preparation and Ground-Truth Grading Details}
\label{app:data_preparation}

\subsection{Preparing $(\question, \just, \ans)$ per Task}

All models are prompted using their standard chat format without any task-specific system prompt engineering.
Each response is expected to contain a free-form justification followed by a final answer, constituting the $(\just, \ans)$ pair.
For models that expose internal reasoning (e.g., extended thinking), only the content surfaced to the user is used as $\just$; any internal scratchpad is excluded, consistent with the definition in Section~\ref{sec:preliminaries}.
The final answer $\ans$ is extracted from the response for use in grading, as described below.

\subsection{Ground-Truth Grading Pipeline}

The grading pipeline differs by task type.

\paragraph{Factual knowledge QA (MMLU, MMLU-Pro, TruthfulQA).}
All factual knowledge QA benchmarks use single-stage rule-based grading (\texttt{gt}).
The model's response is parsed via regex to extract the selected letter choice (A--D for MMLU and TruthfulQA; A--J for MMLU-Pro), which is then compared against the provided answer key.
If no letter choice can be extracted, the response is marked incorrect.

\paragraph{Mathematical reasoning (GSM8K, MATH500).}
Mathematical responses are graded via a two-stage pipeline (\texttt{gt\_verified}), since free-form answers may be expressed in varied but equivalent forms.

\textit{Stage 1 --- Rule-based parsing.}
A regex-based parser first attempts to extract a final numerical or algebraic answer from the response.
If the extracted answer matches the gold answer by string normalization (e.g., stripping whitespace, commas, and ``\$'' symbols), $\GT$ is set to $1$ and the response is not passed to Stage~2.

\textit{Stage 2 --- Model-based parsing and grading (gt\_verified).}
Responses for which Stage~1 fails to confirm correctness are re-processed by GPT-5.2 (temperature $0.0$).
The model first parses whether the response is complete and extracts the final answer (parsing prompt), then compares it against the gold answer for mathematical equivalence (comparison prompt).
$\GT$ is set to $1$ if and only if the parsed answer is deemed equivalent to the gold answer.
The full prompts are given in Section~\ref{app:gpt_grading_prompts} below.

\subsection{GPT-5.2 Prompts}
\label{app:gpt_grading_prompts}

\paragraph{Answer parsing prompt.}

\begin{quote}\ttfamily\small
You are analyzing a math response to extract the final answer.

Response: \{response\}

Task: Determine if this response is FINISHED (has a complete final answer) or UNFINISHED (cut off, incomplete, or missing a final answer).

Rules:\\
1. A response is FINISHED if it clearly states a final numerical or mathematical answer.\\
2. A response is UNFINISHED if there is no clear final answer stated.

If FINISHED, extract the final answer exactly as it appears (just the value, no ``The answer is'' prefix).
If UNFINISHED, output ``N/A''.

Output format (exactly one line): FINISHED: \textlangle{}answer\textrangle{} OR UNFINISHED: N/A
\end{quote}

\paragraph{Answer comparison prompt.}

\begin{quote}\ttfamily\small
You are comparing two mathematical answers to determine if they are equivalent.

Parsed answer from response: \{parsed\_answer\}\\
Gold (correct) answer: \{gold\_answer\}

Task: Are these two answers mathematically equivalent?

Rules:\\
1. Consider mathematical equivalence, not string equality.\\
2. Ignore formatting differences (spaces, commas in numbers, etc.).

Reply with ONLY ``Yes'' or ``No''.
\end{quote}

\subsection{LLM-as-a-Judge Configurations and Prompts}
\label{app:judge_prompts}

All three rater models (GPT-4.1-mini, Claude-Haiku-4.5, Gemini-2.5-Flash-Lite) share the same prompts and decoding configuration: temperature $0.0$, maximum 30 output tokens.

\paragraph{Direct mode (single-turn).}
The rater receives a single prompt and replies with Yes or No.

\textit{AO prompt:}
\begin{quote}\ttfamily\small
Question: \{question\}\\
Response: The answer is \{answer\}.\\
Does the response give the correct answer to the question?
Please reply directly with only a SINGLE Yes or No. Output:
\end{quote}

\textit{AJ prompt:}
\begin{quote}\ttfamily\small
Question: \{question\}\\
Response: \{full\_response\}\\
Does the response give the correct answer to the question?
Please reply directly with only a SINGLE Yes or No. Output:
\end{quote}

\paragraph{Thinking mode (two-turn, AO only).}
The AO setting uses a two-turn exchange to elicit a scratchpad before the forced verdict.
The AJ setting uses the same direct prompt as above (thinking mode does not apply to AJ).

\textit{Turn 1 --- reasoning prompt (up to 256 tokens):}
\begin{quote}\ttfamily\small
Question: \{question\}\\
Response: The answer is \{answer\}.\\
Does the response give the correct answer to the question?
Think step by step about whether this answer is correct.
\end{quote}

\textit{Turn 2 --- forced verdict prompt (up to 30 tokens):}
\begin{quote}\ttfamily\small
Based on your reasoning above, is the answer correct?
Reply with only a SINGLE Yes or No. Output:
\end{quote}

The model's Turn~1 output is appended to the conversation as an assistant turn before Turn~2 is issued, so the final Yes/No verdict is conditioned on the full scratchpad.

\section{Human Study and Platform Implementation Details}
\label{app:human_study}

This appendix documents the human study setup and technical implementation used to collect human judgments for AO (Answer-Only) and AJ (Answer+Justification). It is intended to support reproducibility and auditing.

\subsection{Study Overview (Purpose and Design)}
\textbf{Objective.} The human study evaluates whether human verifiability patterns under AO vs.\ AJ align with those obtained from LLM-as-a-judge, and provides human-grounded measurements for AO/AJ correctness judgments and related subjective signals (e.g., confidence, helpfulness).

\textbf{Key design constraints.}
\begin{itemize}
  \item Each participant completes a \emph{mixed} session containing both AO and AJ items.
  \item No participant sees the same math item more than once (i.e., an item never appears in both AO and AJ for the same participant).
  \item Each math item is targeted to appear at least 3 times in AO and at least 3 times in AJ across all collected sessions.
  \item AO/AJ items are interleaved to reduce systematic position effects (Section~\ref{app:mix-platform:templates}).
\end{itemize}

\subsection{Participant Recruitment and Screening}
Participants were openly recruited through online platforms as well as internal university channels. To ensure that participants met the minimum mathematical reasoning and English comprehension requirements for the main study, we applied a pre-screening test before enrollment. The pre-test consisted of six items in total: five MATH500 problems and one GSM8K problem. For each item, participants were shown the problem together with an LLM-generated \texttt{Answer+Justification}, and were asked to determine whether the LLM's response was correct and, if incorrect, to identify the step at which the error occurred. Only participants who answered at least \textbf{4 out of 5} MATH500 items correctly and the \textbf{1 GSM8K} item correctly were admitted to the main experiment. All participants in the final study were university students. To protect participant privacy, we did not collect additional personal identifying information beyond what was necessary to administer the study.

\subsection{Platform Overview}
We collect data using an in-house online annotation system. The platform is implemented with a React~19 + TypeScript frontend and a Node.js/Express backend. Data are stored in a MySQL relational database. Participants enter the study via a unique invitation link and complete the session entirely in-browser.

\subsection{Item Bank}
The study uses item bank, which contains \textbf{41 items}:
\begin{itemize}
  \item \textbf{40 MATH500 responses} stratified into four categories (TP/TN/FP/FN), \textbf{10 per category}. Each item includes a math question, a model response (justification), and an extracted final answer.
  \item \textbf{1 attention check item} (\texttt{GSM-CHECK}).
\end{itemize}

\subsection{Session Templates}
\label{app:mix-platform:templates}

\subsubsection{Template Structure}
We pre-generate a finite set of fixed \textbf{session templates}. Each template contains \textbf{17 items}:
\begin{itemize}
  \item \textbf{16 math items} sampled from the 40-item MATH500 pool, with \textbf{4 items per category} (TP/TN/FP/FN).
  \item \textbf{1 attention check item} (\texttt{GSM-CHECK}).
\end{itemize}

\subsubsection{AO/AJ Assignment Within a Template}
Within each template, the 16 math items are split evenly:
\begin{itemize}
  \item \textbf{AO (Answer-Only):} show question + extracted final answer only (no justification).
  \item \textbf{AJ (Answer+Justification):} show question + full model justification + extracted final answer.
\end{itemize}

\textbf{Alternation.} The 16 math items are arranged in a strictly alternating pattern:
\[
\texttt{AJ} \rightarrow \texttt{AO} \rightarrow \texttt{AJ} \rightarrow \texttt{AO} \rightarrow \dots
\]
so each template contains exactly \textbf{8 AO} and \textbf{8 AJ} math items.

\textbf{Attention check insertion.} The \texttt{GSM-CHECK} item is always presented in \textbf{AJ} format and inserted at a \textbf{random even index} (0-indexed even positions), i.e., one of positions
\[
0, 2, 4, 6, 8, 10, 12, 14,
\]
which correspond to the 1st, 3rd, 5th, 7th, 9th, 11th, 13th, or 15th item in a 1-indexed display. The insertion does not change the relative alternation pattern among the 16 math items.

\subsubsection{Cross-Template Coverage Constraints and ``Compensation'' Scheduling}
We generate \textbf{20 templates} in total (T1--T20), where T1--T15 are the initial batch and T16--T20 are a supplemental batch.

\textbf{Coverage target.} Across all templates used in data collection, each of the 40 math items is targeted to appear at least:
\[
\texttt{AO} \ge 3 \quad \text{and} \quad \texttt{AJ} \ge 3.
\]

\textbf{Supplemental templates (T16--T20).} The supplemental batch is constructed using a \emph{priority compensation} algorithm:
\begin{itemize}
  \item If an item is under-covered in AO, it is preferentially assigned to an AO slot.
  \item If an item is under-covered in AJ, it is preferentially assigned to an AJ slot.
\end{itemize}
This process continues until all items reach the AO$\ge$3 and AJ$\ge$3 targets, subject to template constraints (16 math items per template; 4 per category; AO/AJ alternation).

\subsubsection{Participant-to-Template Assignment}
Each participant is assigned to exactly one template, recorded in the database field \texttt{mixTemplateId}. A given template is assigned to at most one participant (i.e., templates are not reused), \emph{unless} a session is released/reset due to termination or invalidation.

\subsection{Study Flow and Timing}
Participants proceed through the following states:
\begin{enumerate}
  \item \textbf{Consent.} {\itshape [Welcome, and thank you for considering participation in this study.

The purpose of this study is to understand how everyday users determine whether a proposed answer to a math problem is correct based on the information provided. The information provided will include either the AI's answer to the math question alone, or the AI's answer together with its explanation. During the study, you will see a series of math questions with proposed answers; in some cases, you will also see a justification.

Your task is to decide whether the proposed answer is correct or incorrect using only the information provided in this study.

What You Will Do

Read each math question and its justification, if provided.
For each question, determine whether the provided answer is correct or incorrect.
For some questions, rate how helpful the provided justification was in making your decision.
Time Commitment

The study is expected to take approximately 45–50 minutes.
Risks or Discomforts

We do not expect any notable risks.
This study involves only problem-solving or reading-based judgment tasks.
Voluntary Participation

Your participation is completely voluntary.
You may stop participating at any time without penalty.
If you choose to stop, you may simply close the page.
Privacy and Data Use

We will record your responses and response times.
We will not collect any personally identifying information as part of this research.
Participant Requirements

Do not use scratch paper, calculators, search engines, chatbots, or any other external tools. Rely only on the content provided in this study.
Complete the task using your own judgment only.
Do not switch tabs, take screenshots, or copy content. Any violations will automatically terminate the session]}
  \item \textbf{Instructions.} {\itshape[You will verify whether proposed answers to math problems are correct. You will review 16 math questions with 3 minutes per question. Each question has the AI's proposed answer. Your task is to determine whether each proposed answer is correct or incorrect.]}
  \item \textbf{Active.} Participants answer all 17 items.
  \item \textbf{Completed.} Participants submit and exit the study.
\end{enumerate}

\subsubsection{Per-Item Interaction (Two-Stage)}
Each item is answered in two consecutive stages.

\paragraph{Stage 1: Correctness judgment.}
Participants view the item content (AO or AJ) and respond:
\begin{quote}
\textbf{Correct / Incorrect} (whether the model's proposed answer is correct)
\end{quote}
Timing:
\begin{itemize}
  \item Soft limit: \textbf{3 minutes}. After 3 minutes, the timer turns red and displays a warning.
  \item Hard limit: \textbf{4 minutes}. At 4 minutes the platform auto-submits Stage~1, sets \texttt{timedOut=true}, and records the latest selection if present (otherwise \texttt{null}).
\end{itemize}

\paragraph{Stage 2: Subjective ratings.}
After Stage~1 is locked, participants provide ratings (all required):
\begin{itemize}
  \item \textbf{Confidence} (all settings): 5-point scale (1 = Very Uncertain, 5 = Very Confident).
  \item \textbf{Helpfulness} (AJ only): 5-point scale (1 = Very Unhelpful, 5 = Very Helpful).
\end{itemize}
Timing:
\begin{itemize}
  \item Soft limit: \textbf{60 seconds}.
  \item Hard limit: \textbf{90 seconds} (auto-submit).
\end{itemize}

\subsubsection{ Compensation}
\textbf{Participant-facing compensation text.}

\textit{[Participants were informed that the study used a tiered compensation scheme, with a maximum payment of RMB~100. Completing the study and meeting the basic response requirements guaranteed a base payment of RMB~60. Additional compensation was described as being primarily based on overall judgment accuracy; participants were told that those who responded carefully throughout the study, achieved relatively high accuracy, and completed the subjective ratings conscientiously could receive higher compensation, with average total payment around RMB~70 and a maximum of RMB~100. Participants were also informed that failure to participate seriously---for example, failing the attention check, using external tools, or providing responses that clearly did not meet the task requirements---could result in no compensation. The instruction page further reminded participants to make judgments independently based only on the information shown on the screen, to avoid submitting answers excessively quickly, especially in AO items without justifications, and to complete all subjective ratings carefully, as these were also considered in the overall quality assessment.]}\\

\textbf{Tiering policy}

Internally, participant compensation was tied strongly to judgment accuracy: participants with higher overall accuracy received higher pay, all else being equal. In addition, moderate violations that did not trigger automatic session termination (i.e., fewer than three severe violations) could still reduce compensation. Compensation could also be lowered for behavioral patterns indicative of low-quality participation, including excessively short response times and mechanical subjective ratings with little or no variation across items. Thus, final payment was determined by a combination of overall judgment accuracy, compliance with study rules, and response-quality signals collected during the session.

\subsection{Attention Check}
The session includes one attention check item (\texttt{GSM-CHECK}). The model's response contains an obvious error, and the correct meta-judgment is that the model's answer is incorrect (i.e., the participant should respond \textbf{Incorrect}).

\begin{itemize}
  \item Passing criterion: participant selects \textbf{Incorrect}.
  \item Recorded as: \texttt{participant\_sessions.passedAttentionCheck}.
  \item Presentation: always in \textbf{AJ} format (full justification visible).
  \item Position: inserted at a random even index; across templates, the position distribution spans the set \{1st, 3rd, 5th, 7th, 9th, 11th, 13th, 15th\}.
\end{itemize}

\subsection{Data Integrity and Anti-Cheating Mechanisms}
\subsubsection{Violation Monitoring}
The platform logs potential violations in the \texttt{violation\_events} table. Each event type is assigned a severity level:

\begin{table}[h]
\centering
\small
\begin{tabular}{lll}
\hline
\textbf{Event type} & \textbf{Description} & \textbf{Severity} \\
\hline
\texttt{tab\_switch} & Switch to another browser tab & Severe \\
\texttt{visibility\_hidden} & Window minimized/hidden & Severe \\
\texttt{screenshot\_attempt} & Screenshot attempt detected & Severe \\
\texttt{window\_blur} & Window loses focus & Moderate \\
\texttt{copy\_attempt} & Copy action detected & Moderate \\
\texttt{paste\_attempt} & Paste action detected & Moderate \\
\texttt{right\_click} & Right-click menu opened & Moderate \\
\texttt{devtools\_open} & Developer tools opened & Moderate \\
\hline
\end{tabular}
\caption{Violation events monitored by the platform.}
\end{table}

\textbf{Auto-termination rule.} If a participant accumulates \textbf{3 or more severe violations}
(\texttt{tab\_switch}, \texttt{visibility\_hidden}, \texttt{screenshot\_attempt}),
the platform automatically terminates the session (\texttt{status="terminated"}).

\subsubsection{Session Reset and Template Reuse}
If a session is terminated or otherwise invalidated, the corresponding template may be released and reassigned to a new participant (Section~\ref{app:mix-platform:templates}).

\subsection{Data Logging Schema}
Per-item responses are stored in \texttt{item\_responses}. Key fields include:

\begin{table}[h]
\centering
\small
\begin{tabular}{ll}
\hline
\textbf{Field} & \textbf{Meaning} \\
\hline
\texttt{participantId} & Unique participant identifier \\
\texttt{itemId} & Item ID (e.g., TP01, FN05, GSM-CHECK) \\
\texttt{category} & TP/TN/FP/FN/GSM-CHECK \\
\texttt{condition} & AO or AJ \\
\texttt{questionIndex} & Item position in the sequence (0-indexed) \\
\texttt{responseCorrect} & Stage~1 judgment (true/false/null) \\
\texttt{rtSeconds} & Stage~1 response time (seconds) \\
\texttt{timedOut} & Whether Stage~1 auto-submitted due to timeout \\
\texttt{helpfulness} & Stage~2 helpfulness (AJ only; 1--5) \\
\texttt{confidenceRating} & Stage~2 confidence (all; 1--5) \\
\texttt{confidenceRtSeconds} & Stage~2 response time (seconds) \\
\texttt{submittedAt} & Submission timestamp \\
\hline
\end{tabular}
\caption{Core fields recorded for each item response.}
\end{table}

Session-level metadata (e.g., attention check status, termination status) are stored in \texttt{participant\_sessions}.

\subsection{Inter-Rater Agreement Results}
\label{app:inter_rater_results}

Table~\ref{tab:human_study_full} reports correctness accuracy against ground truth and LLM--human Cohen's~$\kappa$ for each LLM rater individually across all four evaluation settings, together with their average and the human baseline.
$\kappa$ is computed over all (LLM, human) judgment pairs pooled across the 40 items.
Human judgments are available only for AO and AJ settings; $\kappa$ is therefore undefined for the human row.

\begin{table}[h]
\centering
\footnotesize
\setlength{\tabcolsep}{4pt}
\begin{tabular}{l cccc cccc}
\toprule
& \multicolumn{4}{c}{\textbf{Cohen's $\kappa$}} & \multicolumn{4}{c}{\textbf{Accuracy}} \\
\cmidrule(lr){2-5} \cmidrule(lr){6-9}
\textbf{Rater} & \textbf{AO} & \textbf{AO-CoT} & \textbf{AJ} & \textbf{AJ-CoT} & \textbf{AO} & \textbf{AO-CoT} & \textbf{AJ} & \textbf{AJ-CoT} \\
\midrule
\texttt{claude-haiku-4-5}  & $-0.060$ & $0.515$ & $0.529$ & $0.484$ & $0.475$ & $0.900$ & $0.795$ & $0.925$ \\
\texttt{gemini-2.5-flash-lite}      & $0.217$  & $0.436$ & $0.535$ & $0.479$ & $0.650$ & $0.775$ & $0.875$ & $0.875$ \\
\texttt{gpt-4.1-mini}    & $0.039$  & $0.491$ & $0.438$ & $0.501$ & $0.525$ & $0.875$ & $0.875$ & $0.925$ \\
\midrule
LLM (avg.)                          & $0.065$  & $0.481$ & $0.501$ & $0.488$ & $0.550$ & $0.850$ & $0.848$ & $0.908$ \\
\midrule
Human                               & ---      & ---     & ---     & ---     & $0.836$ & ---     & $0.809$ & ---     \\
\bottomrule
\end{tabular}
\caption{Per-rater Cohen's~$\kappa$ (vs.\ human judgments) and ground-truth accuracy across all four evaluation settings on the 40-item human study benchmark. Human judgments are collected only under the AO and AJ settings; $\kappa$ is not applicable for the human row.}
\label{tab:human_study_full}
\end{table}

\subsection{A.8 Collection Status (Snapshot)}
At the time of reporting, data collection status is summarized below:

\begin{table}[h]
\centering
\small
\begin{tabular}{lr}
\hline
\textbf{Metric} & \textbf{Value} \\
\hline
Completed participants & 19 \\
Mean completion time (minutes) & 27.2 \\
Completion time range (minutes) & 16.4 -- 40.4 \\
Attention check pass rate & 100\% (19/19) \\
Participants with 0 violations & 12 (63\%) \\
Participants with $\ge$1 violation & 7 (37\%) \\
Total answered items (incl.\ GSM-CHECK) & 306 \\
\hline
\end{tabular}
\caption{Data collection snapshot.}
\end{table}

\end{document}